 \documentstyle{mn}
\input psfig

\newif\ifAMStwofonts

\newcommand{\figstart}[1]  {\begin{figure} \psfig{#1}}
\newcommand{\lfigstart}[1] {\begin{figure*} \psfig{#1}}
\newcommand{\figend} {\end{figure}}
\newcommand{\lfigend} {\end{figure*}}

\newcommand{\kms} {$\mbox{km.s}^{-1}$}
\newcommand{\Msun} {$\mbox{M}_{\sun}$}

\newcommand{\magsec} {$\mbox{mag.arcsec}^{-2}$} 
\newcommand{\Ag} {\AA~} 
\newcommand{\diff} {\mbox{d}}

\def\spose#1{\hbox to 0pt{#1\hss}}
\def\lta{\mathrel{\spose{\lower 3pt\hbox{$\sim$}}
    \raise 2.0pt\hbox{$<$}}}
\def\gta{\mathrel{\spose{\lower 3pt\hbox{$\sim$}}
    \raise 2.0pt\hbox{$>$}}}
\def\etal{et al.~}
\def\MBH{{M_{\rm BH}}}

\def\spose#1{\hbox to 0pt{#1\hss}}
\def\lta{\mathrel{\spose{\lower 3pt\hbox{$\sim$}}
    \raise 2.0pt\hbox{$<$}}}
\def\gta{\mathrel{\spose{\lower 3pt\hbox{$\sim$}}
    \raise 2.0pt\hbox{$>$}}}
\newdimen\hssize
\newdimen\hdsize
\ifoldfss
  \ifCUPmtlplainloaded \else
    \NewTextAlphabet{textbfit} {cmbxti10} {}
    \NewTextAlphabet{textbfss} {cmssbx10} {}
    \NewMathAlphabet{mathbfit} {cmbxti10} {} 
    \NewMathAlphabet{mathbfss} {cmssbx10} {} 
  \fi
  \ifAMStwofonts
    \ifCUPmtlplainloaded \else
      \NewSymbolFont{upmath} {eurm10}
      \NewSymbolFont{AMSa} {msam10}
      \NewMathSymbol{\upi}     {0}{upmath}{19}
      \NewMathSymbol{\umu}     {0}{upmath}{16}
      \NewMathSymbol{\upartial}{0}{upmath}{40}
      \NewMathSymbol{\leqslant}{3}{AMSa}{36}
      \NewMathSymbol{\geqslant}{3}{AMSa}{3E}

       \let\ge=\geqslant
    \fi
  \fi
\fi 

\ifnfssone
  \newmathalphabet{\mathit}
  \addtoversion{normal}{\mathit}{cmr}{m}{it}
  \addtoversion{bold}{\mathit}{cmr}{bx}{it}
  \newmathalphabet{\mathbfit} 
  \addtoversion{normal}{\mathbfit}{cmr}{bx}{it}
  \addtoversion{bold}{\mathbfit}{cmr}{bx}{it}
  \newmathalphabet{\mathbfss} 
  \addtoversion{normal}{\mathbfss}{cmss}{bx}{n}
  \addtoversion{bold}{\mathbfss}{cmss}{bx}{n}
  \ifAMStwofonts
    \ifCUPmtlplainloaded \else
      \UseAMStwoboldmath
      \makeatletter
      \new@mathgroup\upmath@group
      \define@mathgroup\mv@normal\upmath@group{eur}{m}{n}
      \define@mathgroup\mv@bold\upmath@group{eur}{b}{n}
      \edef\UPM{\hexnumber\upmath@group}
      \new@mathgroup\amsa@group
      \define@mathgroup\mv@normal\amsa@group{msa}{m}{n}
      \define@mathgroup\mv@bold\amsa@group{msa}{m}{n}
      \edef\AMSa{\hexnumber\amsa@group}
      \makeatother
      \mathchardef\upi="0\UPM19
      \mathchardef\umu="0\UPM16
      \mathchardef\upartial="0\UPM40
      \mathchardef\leqslant="3\AMSa36
      \mathchardef\geqslant="3\AMSa3E

       \let\ge=\geqslant
    \fi
  \fi
\fi 
\ifnfsstwo
  \DeclareMathAlphabet{\mathbfit}{OT1}{cmr}{bx}{it}
  \SetMathAlphabet\mathbfit{bold}{OT1}{cmr}{bx}{it}
  \DeclareMathAlphabet{\mathbfss}{OT1}{cmss}{bx}{n}
  \SetMathAlphabet\mathbfss{bold}{OT1}{cmss}{bx}{n}
  \ifAMStwofonts
    \ifCUPmtlplainloaded \else
      \DeclareSymbolFont{UPM}{U}{eur}{m}{n}
      \SetSymbolFont{UPM}{bold}{U}{eur}{b}{n}
      \DeclareSymbolFont{AMSa}{U}{msa}{m}{n}
      \DeclareMathSymbol{\upi}{0}{UPM}{"19}
      \DeclareMathSymbol{\umu}{0}{UPM}{"16}
      \DeclareMathSymbol{\upartial}{0}{UPM}{"40}
      \DeclareMathSymbol{\leqslant}{3}{AMSa}{"36}
      \DeclareMathSymbol{\geqslant}{3}{AMSa}{"3E}

       \let\ge=\geqslant
    \fi
  \fi
\fi 

\ifCUPmtlplainloaded \else
  \ifAMStwofonts \else 
    \def\upi{\pi}
    \def\umu{\mu}
    \def\upartial{\partial}
  \fi
\fi

\title[Dynamical models of NGC~3115]
{Dynamical models of NGC~3115\thanks{Based on observations taken with the Canada-France-Hawaii
Telescope, operated by the National Research Council of Canada, the Centre National de la
Recherche Scientifique of France, and the University of Hawaii}}
\author[Emsellem, Dejonghe \& Bacon]
{Eric Emsellem$^{1,2}$,
Herwig Dejonghe$^3$, and Roland Bacon$^2$ \\
$^1$ European Southern Observatory, Karl-Schwarzschild Strasse 2, 85748
Garching b. M\"unchen, Germany \\
$^2$ Centre de Recherche Astronomique de Lyon, Observatoire de Lyon, 9 av. Charles-Andr\'e,
69561 Saint-Genis Laval Cedex, France \\
$^3$ Sterrenkundig Observatorium, Universiteit Gent, Krijgslaan 281,
B9000 Gent, Belgium}

\date{Accepted .
      Received}

\pagerange{\pageref{firstpage}--\pageref{lastpage}}
\pubyear{1998}

\begin{document}

\maketitle

\label{firstpage}

\begin{abstract}
We present new dynamical models of the S0 galaxy
N3115, making use of the available published photometry and kinematics
as well as of two-dimensional TIGER spectrography. 

The models are based on a detailed model of the luminosity
distribution built using an MGE fit on HST/WFPC2 and ground-based
photometric data. We first examined the kinematics in the central
$40\arcsec$ in the light of two-integral $f(E, J)$ models.  Jeans
equations were used to constrain the mass to light ratio, and the
central dark mass whose existence was suggested by previous studies.
The even part of the distribution function was then retrieved via the
Hunter \& Qian formalism. We thus confirmed that the velocity and
dispersion profiles in the central region could be well fit with a
two-integral model, given the presence of a central dark mass of $\sim
10^9$~\Msun.  However, no two-integral model could fit the $h_3$
profile around a radius of about $25\arcsec$ where the outer disc
dominates the surface brightness distribution.

Three integral analytical models were therefore built using a
Quadratic Programming technique. These models showed that three
integral components do indeed provide a reasonable fit to the
kinematics, including the higher Gauss-Hermite moments. Again, models
without a central dark mass failed to reproduce the observed
kinematics in the central arcseconds. This clearly supports the
presence of a nuclear black hole of at least $6.5\times10^8$~\Msun in
the centre of NGC~3115. These models were finally used to estimate the
importance of the dark matter in the outer part of NGC~3115, suggested
by the flat stellar rotation curve observed by Capaccioli et al.
(1993).

This study finally points out the difficulty of integrating
independently published data in a coherent and consistent way, thus
demonstrating the importance of taking into account the details of the
instrumental setup and the reduction processes.

\end{abstract}

\section{Introduction}

Early-type disc galaxies pose a specific problem for dynamical
modeling as neither the bulge nor the disc can be neglected in the
computation of the gravitational potential. First, it is generally
difficult to disentangle the photometric components according to a
priori fixed criteria such as the ellipticity or surface brightness
profiles. Although there exists different methods to build
corresponding realistic photometric models, they are usually limited
to two component systems (e.g. disc plus spheroid, Byun \& Freeman
1995) which do often not reflect the true morphology of the galaxy.
Then, the different internal dynamics of each component has to be
taken into account. Asymetrical Line Of Sight Velocity Distributions
(LOSVDs) observed in discy ellipticals (e.g. Scorza \& Bender 1995)
confirmed the fact that the classically measured velocity $V$ and
dispersion $\sigma$ do not sufficiently constrain their dynamical
structure.  Because of these difficulties, the analysis is usually
restricted to the central region, or to regions where one component
dominates the light (e.g. Jarvis \& Freeman 1985).

A number of techniques to build realistic distribution
functions of such complex objects have recently been 
developed. Hunter \& Qian (1993) proposed a viable scheme
to retrieve the even part of the axisymmetric
two integral distribution function
corresponding to a given mass model. It was used by
van den Bosch \& de Zeeuw (1996) for a set of models including a
spheroid and a disc. In principle, this method can be
applied to more complex potentials: this will be demonstrated in the
present paper. As a further step, three integral axisymmetric
models can be built, using for example the
Schwarzschild method in which one derives the best combination
of orbits constrained to fit a set of observational data
(e.g. van der Marel et al. 1998, Cretton \& van den Bosch 1998). 
In this paper, we wish to present the application of another
technique, namely Quadratic Programming (Dejonghe 1989)

NGC~3115 is interesting in many different aspects. First it has long
been assumed to be the prototype of the S0 galaxy type: a bulge
dominated galaxy with an embedded disc and very little gas and dust.
With a distance of about 10~Mpc and a nearly edge-on inclination,
NGC~3115 is also a perfect target for optical observations. Numerous
studies were focused on the understanding of its morphology and
kinematics. NGC~3115 contains a double disc structure with an outer
Freeman type II disc extending up to $\sim 140\arcsec$ and a nuclear
disc (inside $3\arcsec$) clearly revealed by the HST/WFPC2 pictures
(Kormendy \etal 1996). Its outer disc was found to exhibit a weak
spiral arm structure (Capaccioli \etal 1987, Silva \etal 1988) and to
thicken outwards (Capaccioli \etal 1988).  The spheroidal component
does not seem to follow the classical $r^{1/4}$ law, and the flattened
halo of NGC~3115 extends up to $20\arcmin$ (Capaccioli \etal 1987).

The kinematics of NGC~3115 also deserved a complete survey:
Illingworth \& Schechter (1982) presented an extensive series of
spectroscopic observations demonstrating that the bulge of this galaxy
was nearly consistent with being an isotropic rotator. NGC~3115 has
also been one of the best candidates for the presence of a central dark
mass since the spectroscopic observations of Kormendy \& Richstone
(1992): spherical models suggested an increase of $M/L$ by more than a
factor of 10 inside $2\arcsec$, which is difficult to account for with normal
stellar populations.  This conclusion was based on the observed
velocity gradient and central gaussian velocity dispersion $\sigma_0$,
the latter nearly reaching 300~\kms at a resolution of $\sim 1\arcsec$
FWHM. This trend has been very recently confirmed with FOS
spectroscopy (Kormendy \etal 1996), yielding $\sigma_0 \sim 445$~\kms at
HST resolution. Other studies mainly focused on the major-axis
kinematics (Carter \& Jenkins 1993, van der Marel \etal 1994, Bender
\etal 1994, Fisher 1997). Finally Capaccioli \etal (1993) have shown
that the stellar rotation and dispersion profiles were flat at large
radii ($R > 100\arcsec$), thus indicating the presence of a massive
dark halo.

In this paper, our aim is to make detailed models of this early type
disc galaxy, e.g. constrain its distribution function, 
fully taking into account the available published data. 
This includes, for the first time, two-dimensional
spectroscopic observations. In Sect.~2 we describe the photometry and spectroscopic
observations used for the modeling, including new HRCAM images, the
WFPC2 $V$ band image as well as original TIGER spectrography of the
central region. The photometric models based on the MGE formalism
are presented in Sect.~3 and their corresponding dynamical Jeans
models in Sect.~4.  Numerical two integral distribution functions are
derived using the Hunter \& Qian method and analysed in Sect.~5.  Two
and three integral models are computed in Sect.~6 with the help of the
Quadratic Programming method. Conclusions are drawn in Sect.~7.

\section{Observational data}
\label{sec:obs}

\subsection{Photometry}

Dynamical modeling first implies the availability of a luminosity
model. This requires images with high resolution to get details on the
central region, but also with a large field of view to correctly
sample the line-of-sight at large radii.  A wide field $V$ band image,
obtained with a focal reducer on the 1.23m telescope at Calar Alto,
was kindly put at our disposal by Cecilia Scorza. We then retrieved
the $V$ band WFPC2 image (F555W) of NGC~3115 from the HST archives at ECF
(PI Faber). We finally used a $V$ band image obtained by Jean-Luc
Nieto in April 1992 with HRCAM at the CFHT. All these images were
reduced and normalized using aperture photometry available in the
literature (Poulain 1986 \& 1988, de Vaucouleurs \& Longo 1988). These
calibrations led to individual errors smaller than 0.05~\magsec with
an excellent overall agreement.  The characteristics of the different
images are given in Table~\ref{tab:photo}.
\begin{table}
\caption[]{Characteristics of the $V$ band images of NGC~3115.}
\begin{center}
\begin{tabular}{llll}
\hline
& Calar Alto & HRCAM & WFPC2 \\
pixel size & $1\farcs55$ & $0\farcs11$ & $0\farcs0455$ \\
seeing ($\sigma_{\star}$) & $\sim 3 \arcsec$ & $0\farcs7$ &  - \\
field of view & $10\arcmin\times15\arcmin$ & $70\arcsec \times 110\arcsec$ & 
$34\arcsec \times 33\arcsec$ \\
exp. time & 240s & 120s & 1030s \\
airmass & 1.65 & 1.13 & -- \\
\hline
\end{tabular}
\end{center}
\label{tab:photo}
\end{table}

\subsection{Spectroscopy and kinematics}

\subsubsection{TIGER data}
\label{sec:tigdata}

\lfigstart{figure=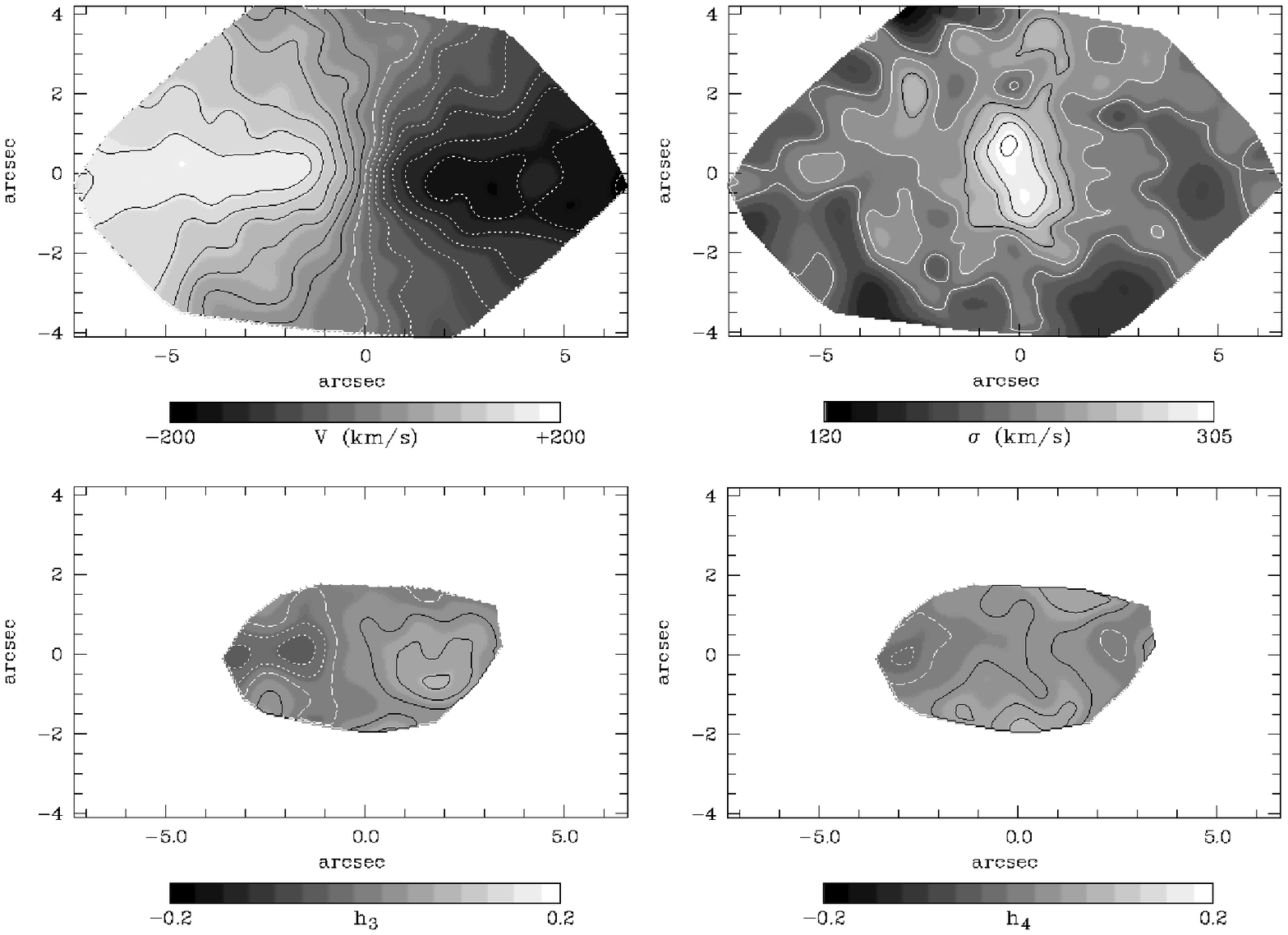,width=17.6cm}
\caption[]{TIGER stellar kinematical maps: velocity (top left), 
dispersion (top right), $h_3$ (bottom left) and $h_4$ (bottom right).
The isocontour step is 25~\kms for both $V$ and $\sigma$, and 0.05 for $h_3$ and
$h_4$.}
\label{fig:tigmaps}
\lfigend

\lfigstart{figure=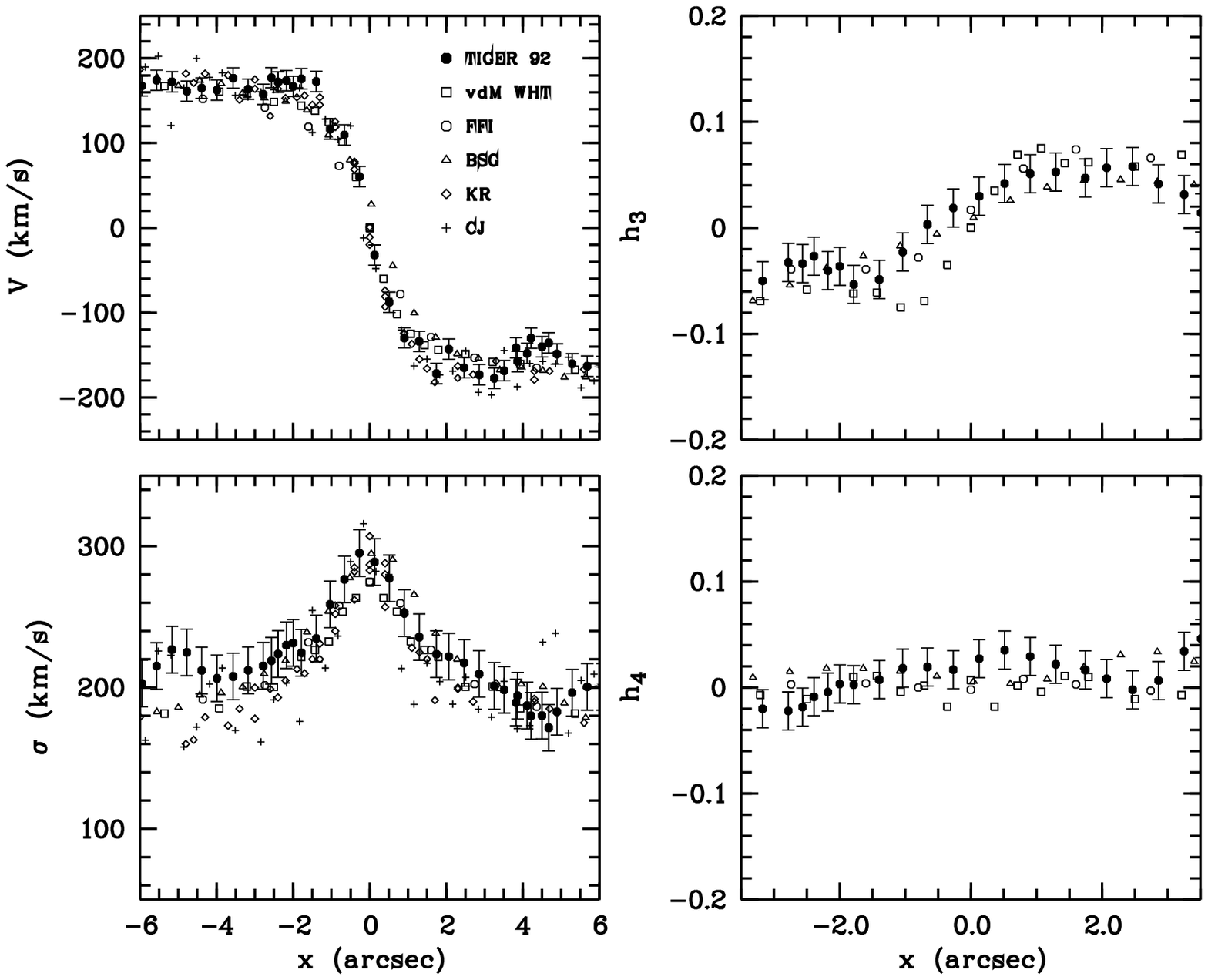,width=17.6cm}
\caption[]{Comparison between the different published kinematics and the
TIGER 1992 data. Each symbol corresponds to a reference as shown in the $V$ panel.
F97 = Fisher 1997; vdM = van der Marel \etal 1994 (WHT data only); BSG = Bender \etal 1994;
KR = Kormendy \& Richstone 1992 (exposures 29F48, 29F42, 34F16; see KR92); 
CJ = Carter \& Jenkins 1993.}
\label{fig:compkin}
\lfigend

We observed NGC~3115 with the TIGER integral field spectrograph (Bacon \etal 1995) at the CFHT in April
1992 and subsequently in April 1996. The lens diameter was set to
$0\farcs39$ and the spectral domain to 5130--5520~\AA\ for both runs.
The spectra were extracted and calibrated in the standard way using
the TIGER software written at the Lyon Observatory.  This includes
bias subtraction, flat-fielding, spectra extraction, wavelength
calibration, cosmic rays removal and differential refraction
correction (see Emsellem \etal 1996 for more details). The individual
exposures were then accurately centred and merged. Variations in the
overall atmospheric transparency were smaller than 1\% and corrected.
We thus obtained one set of spectra for each run, consisting of 624
and 391 spectra for the run92 and run96 respectively. The final seeing
of each data cube was estimated a posteriori by comparing images
reconstructed from the spectra with the $V$ band WFPC2 image. 
Details on the characteristics of both sets of data are
given in Table~\ref{tab:tiger}. Note that although the set of data
obtained in April 1996 has a significantly better spatial and spectral
resolution, we will mainly present the data set obtained in April 1992
as the signal to noise is higher and the field of view larger.
\begin{table}
\caption[]{Observational characteristics of the TIGER spectrographic 
exposures. The fields of view are only indicative as the coverage of
the merged TIGER spectra are not rectangular. All individual frames
had an exposure time of 30 mn.}
\begin{center}
\begin{tabular}{llll}
\hline
& 1992 Run & 1996 Run \\
\hline
Lens diam. & $0\farcs39$ & $0\farcs39$ \\
Field of view    & $14\farcs5 \times 7\farcs9$ & $7\farcs9 \times 7\farcs9$ \\
\# of exposures & 4 & 2 \\
\# of spectra & 624 & 391 \\
Spect. sampling & 1.8 \Ag.$\mbox{pixel}^{-1}$ & 1.5 \Ag.$\mbox{pixel}^{-1}$ \\
Spect. resolution & 3.7 \Ag (FWHM) & 3.2 \Ag (FWHM) \\
Instr. broadening ($\sigma$) & 94 \kms & 78~\kms \\
Spect. domain & 5130 - 5520 \Ag & 5130 - 5520 \Ag \\
Seeing (FWHM) & $1\farcs24$ & $0\farcs8$ \\
Stellar templates & $\eta$ Cygni  & HR 127065 \\
& K0III & K0III \\
\hline
\end{tabular}
\end{center}
\label{tab:tiger}
\end{table}

In order to determine the stellar kinematics we also obtained
exposures of a few stellar templates: mainly K giants for run92 and G,
K and M giants for run96. All spectra were logarithmically rebinned,
the stellar continuum was approximated using a fifth degree polynomial
and subtracted. We then derived the 2D kinematics of the central
region of NGC~3115 using the FCQ algorithm (Bender 1990) and an
optimal template built with the available stellar spectra. This
provides the full LOSVD for each spatial element. The LOSVDs were then
parametrized with the Gauss-Hermite functions.  True moments were also
estimated from the positive part of the parametrized LOSVDs (see
Emsellem \etal 1996). The results were found to be rather insensitive
to the choice of the template, so we decided to use the spectrum of a
single K0 giant for both runs. We finally built the maps for each
kinematical quantity: $V$, $\sigma$, $h_3$ and $h_4$ which correspond
to the Gauss-Hermite parametrization as well as the estimations of the
four first true moments of the LOSVD $\tilde{V}$, $\tilde{\sigma}$,
the skewness and the kurtosis.

The mean velocity, velocity dispersion (gaussian fit), $h_3$ and $h_4$
two-dimensional maps are presented in Fig.~\ref{fig:tigmaps} (run92).
We detect a slight tilt of $10\degr$ of the velocity minor-axis in the
1992 data. The low signal to noise of the new TIGER data set prevented
us to confirm this: it would therefore be useful to obtain new high
spatial resolution kinematical data along the minor-axis. Note that
the $h_3$ and $h_4$ maps only include spectra with a signal to noise
higher than 20.

\subsubsection{Comparison with published kinematics}
\label{sec:pdata}

All kinematical data available were compiled, and the
FOS/HST ($0\farcs21$ square aperture) and SIS/CFHT data obtained by
Kormendy \etal (1996) scanned.  As emphasized by different authors
(e.g. van der Marel \etal 1994) the gaussian velocity $V$ and
dispersion $\sigma$ are not always good estimations of the true
$\tilde{V}$ and $\tilde{\sigma}$ moments. This is particularly
important in the case of NGC~3115 where high values of $h_3$ have been
measured (e.g. van der Marel \etal 1994).  Therefore, we mainly
focused on the data sets including higher order moments of the LOSVDs:
van der Marel \etal (1994, hereafter vdM+94), Bender \etal (1994, hereafter B+94),
Fisher (1997, hereafter F97) and the TIGER data.  In these
cases, we used the Gauss-Hermite expansion of the LOSVDs to estimate
$\tilde{V}$ and $\tilde{\sigma}$ using the same method as used by
Emsellem \etal (1996). Still we made use of other published kinematics
such as Kormendy \& Richstone (1992, hereafter KR92), Carter \&
Jenkins (1993), and the recent high spatial resolution FOS/HST and
SIS/CFHT data (Kormendy \etal 1996, hereafter K+96).  We finally
included the data of Illingworth \& Schechter (1982, hereafter IS82)
for offset axes and of Capaccioli \etal (1993, hereafter C+93) for the
kinematics at large radii along the major-axis. The characteristics of
each data set are given in Table~\ref{tab:kine}.
\begin{table*}
\caption[]{Observational characteristics of the kinematical (long-slit) data used
in this paper (for the FOS/HST data, see Kormendy \etal 1996).}
\begin{center}
\begin{tabular}{lcccc}
\hline
Ref. & slit width & $\sigma_v$ & seeing (FWHM) & $h_3, h_4$ \\
\hline
Illingworth \& Schechter 1982 & $2$--$3\arcsec$ & $100$--$130$~\kms & $1\farcs5$--$5\arcsec$ 
& no \\
Capaccioli \etal 1993 & $1$--$1\farcs6$ & $\sim 40$~\kms & $> 1\farcs5$ 
& no \\ 
Kormendy \& Richstone 1992 & $0\farcs5$ & $68$~\kms & $1\arcsec$--$1\farcs6$ 
& no \\
Carter \& Jenkins 1993 & $0\farcs45$ & $10$~\kms & $0\farcs8$ 
& no \\
van der Marel \etal 1994 & $1\arcsec$ & $38$~\kms & $0\farcs96$ & yes \\
Bender \etal 1994 & $2\farcs1$ & $46$~\kms & $1$--$2\farcs5$ & yes \\
Fisher 1997 & $2\arcsec$ & $75$~\kms & $> 1\farcs5$ & yes \\
Kormendy \etal 1996 & $0\farcs26$ & $\sim 35$~\kms & $0\farcs57$ & - \\
\hline
\end{tabular}
\end{center}
\label{tab:kine}
\end{table*}

In Fig.~\ref{fig:compkin}, we compare the TIGER kinematics along
the major-axis with the ground-based data with similar resolution:
they are all in good agreement. The
error bars of the TIGER data were determined by the statistics on the
fluctuations inside a spatial resolution element, here taken as the
FWHM of the PSF\footnote{Lens to lens fluctuations in the
reconstructed kinematical maps are a good estimator of the errors,
formal and instrumental}.  As it turns out, this defines a beam
containing on average 7 spectra.

\subsection{Convolution and pixel integration}

The comparison between observed kinematics and theoretical models
requires to take into account the observational characteristics.  The
seeing smearing and pixel binning have been computed through a double
quadrature (see Appendix A). All the comparisons presented in the
following sections include this processing, according to the
parameters of the relevant data set as tabulated in
Table~\ref{tab:kine}.  Note that the PSF of the FOS data was derived 
using an approximation as given in van der Marel \etal (1997).

\subsection{Uncertainties in the kinematics}
\label{sec:error}

Additional factors linked with the instrumental setup 
or the data reduction processes can significantly influence the measured
kinematics, and in particular the higher order Gauss-Hermite moments.
Firstly, an error in the centering or position angle of the slit induces
a change in the observed kinematics. In the case
of NGC~3115, an offset of $\Delta \alpha = 2 \degr$ (major-axis) in the
position angle of the slit gives $\Delta h_3 \sim 10$\% at $R = 35\arcsec$, 
while the change in velocity amounts to only about 3\%.
It is also important to control the spectral filtering 
applied for the retrieval of the LOSVDs as it could lead
to underestimate the higher order moments.
Finally, the obtained kinematics may be sensitive to the method
used (e.g. Fourier fitting, FCQ, etc). In the following paragraph,
we discuss such an effect on the central FOS LOSVD.

\subsubsection{The central $h_4$ value}
\label{sec:h4}

The central observed FOS LOSVD is one of the best data set we can use
to constrain the central dark mass in NGC~3115, as it
has the highest available spatial resolution, it includes the higher
order kinematical terms and does not depend on the odd part of the distribution
function (or very weakly since $r = 0\farcs01$). K+96 quotes a central
value of $h_4 = 0.10 \pm 0.03$. However, higher order Gauss-Hermite moments
may significantly depend on the instrumental setup and reduction
procedures.  We thus wish here to assess the robustness of the central
$h_4$ value before attempting any detailed modeling.

We have therefore retrieved and reduced the FOS spectra and tried to
reproduce the analysis mentioned in K+96. Our velocity and dispersion
profiles obtained with the Fourier Quotient method (FQ) are compatible
(within 15~km.s$^{-1}$) with the ones quoted by K+96: e.g.  we find a
value of $\sigma(0) = 448$~km.s$^{-1}$ to be compared with $\sigma(0)
= 443 \pm 18$~km.s$^{-1}$ as given by K+96. However, the shape of the
LOSVD, and thus the dispersion and $h_4$ values, obtained using the
Fourier Correlation Quotient (FCQ) can be severely influenced by
template mismatch, a bad continuum subtraction and the spectral
filtering (see also vdM+94). These would in turn affect the high frequencies
of the LOSVD, thus diminishing its peakedness, and/or the low frequencies
smoothing out its wings. Moreover, large $h_4$ values may be anyway
difficult to obtain, as already emphasized by van der Marel (1994).

We have thus simulated artificial FOS spectra by convolving a FOS
stellar template with different numerical LOSVDs, including the noise
level present in the observed FOS spectra\footnote{This noise does not
appear in Fig.~3 of K+96, so we assumed that the spectra have been
significantly filtered for the plot.} of NGC~3115. We then calculated
the LOSVDs via FCQ, as in K+96, and parametrized them using
Gauss-Hermite functions. These simulations clearly demonstrate that
large $h_4$ values are underestimated (and as a consequence,
gaussian dispersion obtained via FCQ probably overestimated):
the dominant effect is a strong filtering of the central peak,
although the large wings of the LOSVDs are also significantly affected. 
Considering the central LOSVD presented in K+96 and the 
results of our simulations, we can estimate
the true value of the central $h_4$ to be $\sim 0.17$ with a possible
range $0.14 < h_4 < 0.21$. But as discussed above, these values are
ill-constrained and an accurate estimation would require additional
observations. We should finally note that this sensitivity 
is particularly important for the FOS/HST
data: at a lower spatial resolution, gradients are not as extreme and
the LOSVDs closer to Gaussians.

\section{The MGE luminosity model}
\label{sec:MGE}
\figstart{figure=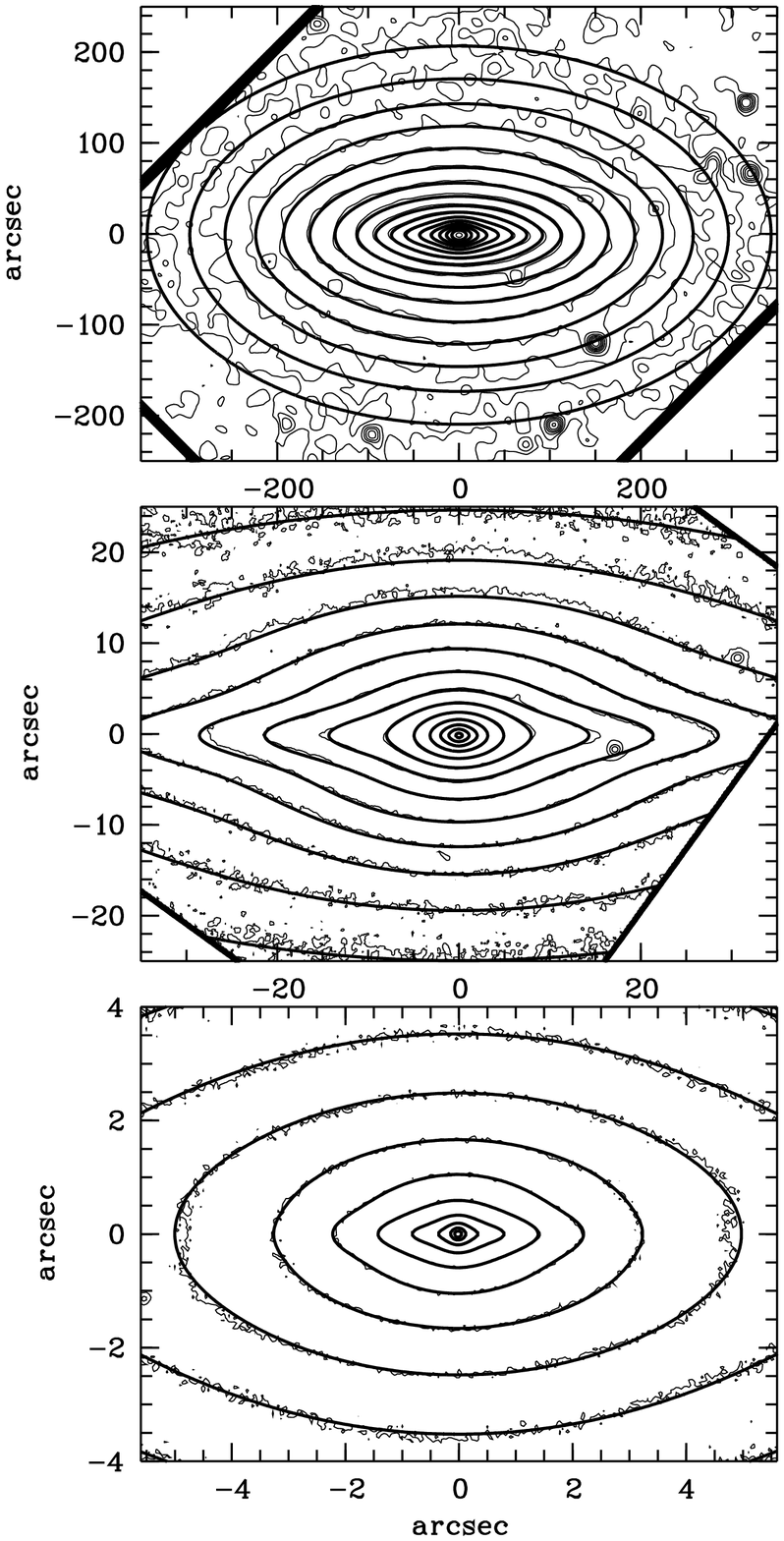,width=\hssize}
\caption[]{Isophotes of the $V$ band images of NGC~3115 (thin lines)
and of the corresponding convolved MGE model (19 components): wide field image (Calar
Alto, top), HRCAM (middle) and WFPC2 (bottom). The isophotes step is
0.5~\magsec, and the faintest isophotes are 25, 20.5 and 17.5~\magsec
respectively }
\label{fig:photo}
\figend
We applied the MGE technique (Monnet \etal 1992, Emsellem \etal 1994, Emsellem 1995)
to the available $V$ band images to build a complete photometric model
of NGC~3115. The individual Point Spread Functions were taken into
account to obtain a deconvolved model which fits the data from the
central HST/WFPC2 pixel ($0\farcs0455$) to $300\arcsec$.  In
Fig.~\ref{fig:photo} we present the isophotes of NGC~3115 at three
different scales compared with the MGE $V$ band model. The fit is
excellent at all scales (see also Fig.\ref{fig:showcusp}). The only significant discrepancies are
observed at $R \sim 5\arcsec$ where the isophotes are boxy due to a detected
spiral-like structure, and at $R \sim 30 \arcsec$ where the
galaxy slightly departs from axisymmetry.  The largest component has a
FWHM of $895\arcsec$ and an axis ratio of 0.85. \\
We did not attempt to go beyond this scale as
the photometry becomes rather uncertain. Hence, at radii larger than
$600\arcsec$ along the major-axis, the MGE model decreases more
rapidly (exponentially) than the $B$ band profile published by
Capaccioli \etal (1987). However we tested that this does not
significantly influence the dynamics up to $150\arcsec$, radius of the
last observed kinematic data point.
\figstart{figure=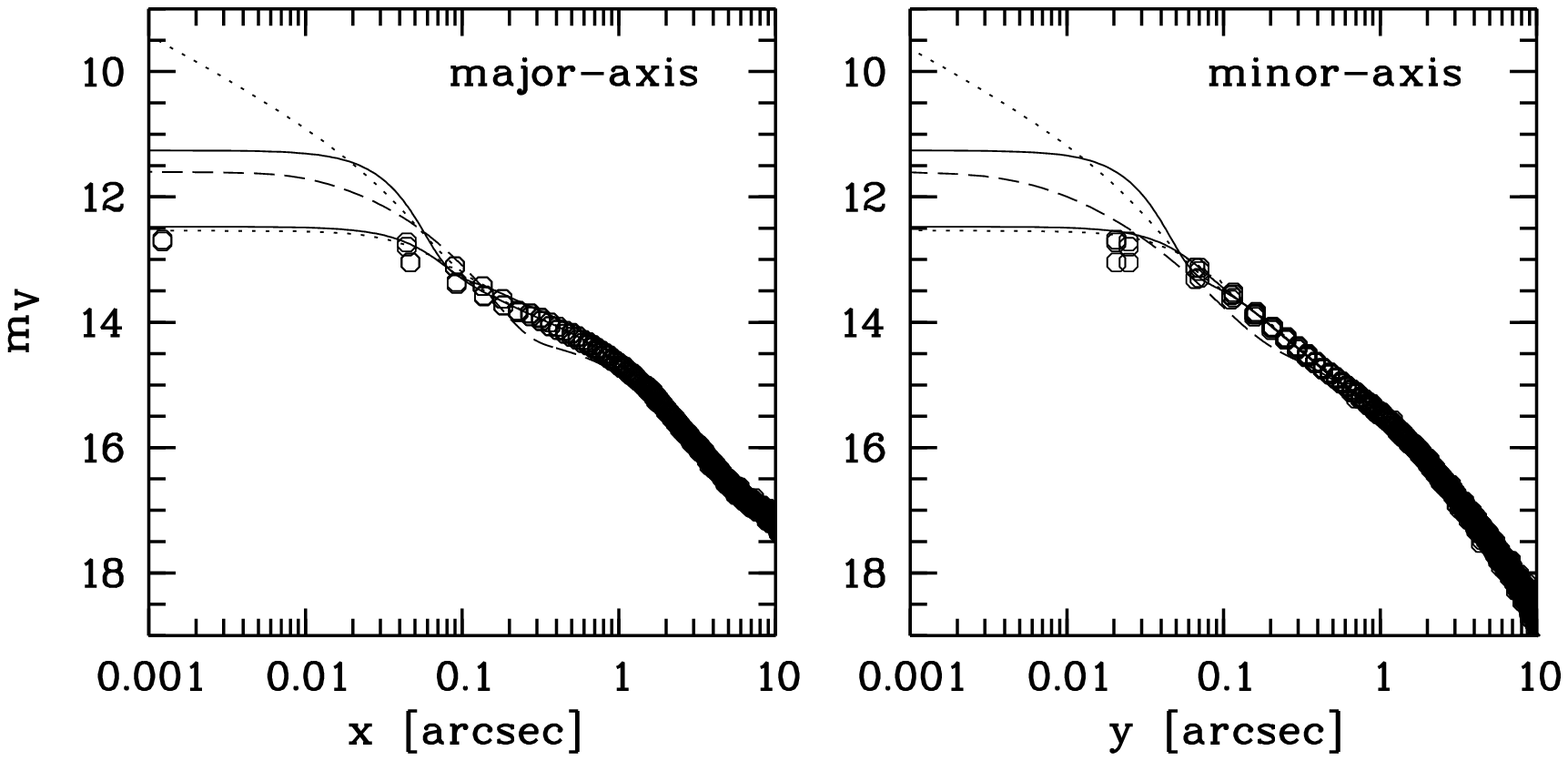,width=\hssize}
\caption[]{Central $V$ band major and minor-axis profiles. The HST/WFPC2 points are 
compared with the MGE models and a three integral model (dashed
curves, see Sect.~\ref{sec:QP}) along the major-axis (left panel) and
minor-axis (right panel). In each panel, the lower solid and dotted curves
correspond to the convolved (and binned) MGE models with a central
gaussian (solid lines) or a cusp ($\gamma = 1.5$, dotted lines; see
Appendix B). The corresponding {\em deconvolved} MGE models are also
shown for comparison (two upper curves in each panel).}
\label{fig:showcusp}
\figend
\figstart{figure=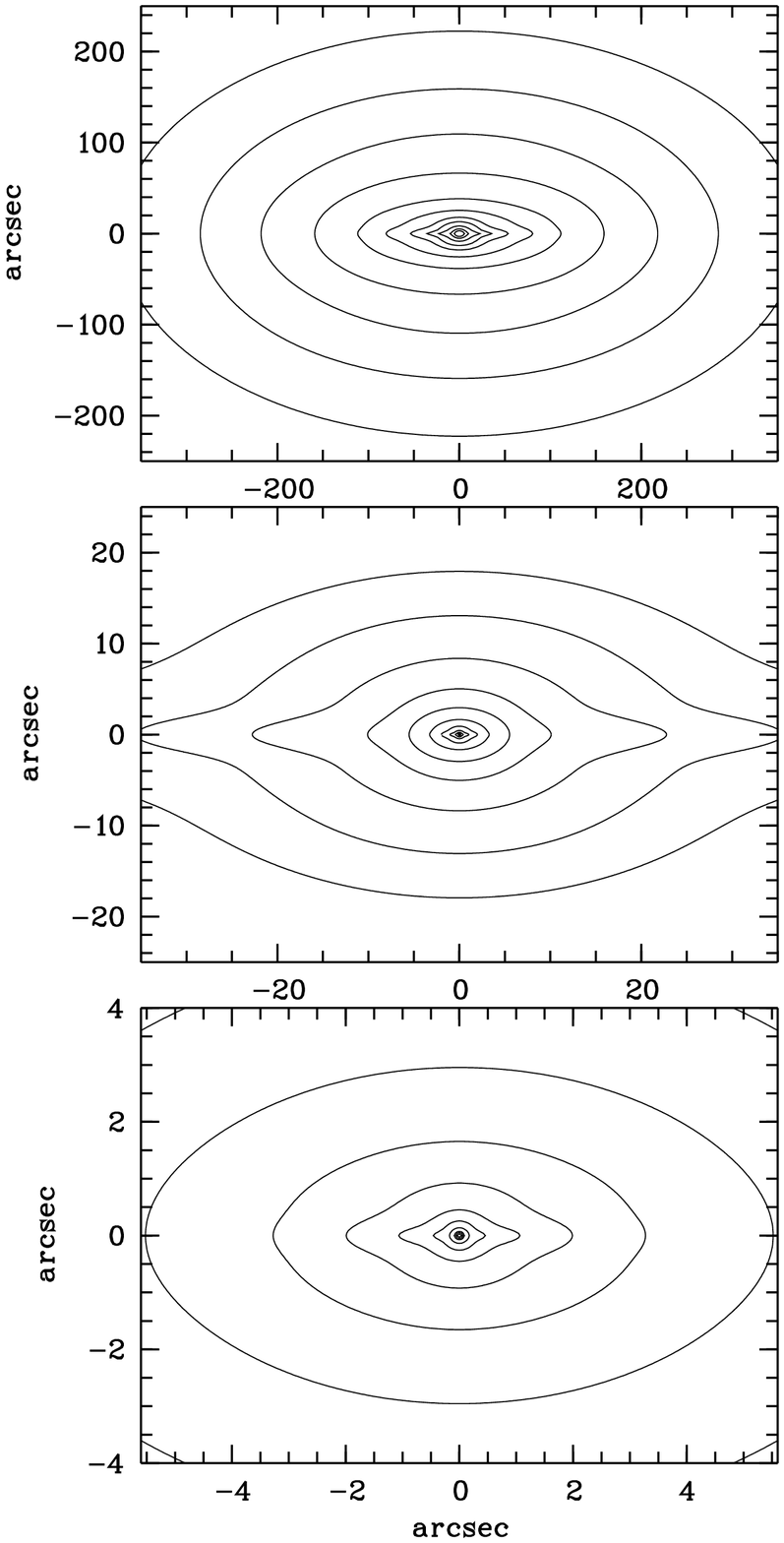,height=15cm}
\caption[]{Isophotes of the deprojected (and deconvolved) MGE model of NGC~3115 ($V$ band) 
in the meridian plane $(R, z)$: the steps are 0.5 in
$\log{\left(\mbox{L$_{\odot}$.pc$^{-3}$}\right)}$, with the faintest
isophotes corresponding to (from top to bottom) -4, -1.5, and 0.5
respectively. The nucleus peaks at $\sim 5.42$ ($\sim
1.29\;10^7$~L$_{\odot}$.pc$^{-3}$, model with a central gaussian).}
\label{fig:spatial}
\figend

In Fig.~\ref{fig:spatial}, we show the isophotes of the deprojected MGE model
for an inclination angle of $86\degr$. The different components clearly appear
in this plot:
\begin{itemize}
\item the flattened extended halo which dominates the light of the galaxy
at radii larger than $100\arcsec$ ($\epsilon \sim 0.65$ at $300\arcsec$).
\item the outer disc which extends up to $\sim 100\arcsec$ and 
is truncated at $R \sim 5\arcsec$ (Freeman type II disc);
\item the inner spheroid which exhibits a boxy shape and
an ellipticity in the range $0.4-0.6$;
\item the inner or nuclear disc which continues up the the centre
and has a radial extent of about $3\arcsec$;
\item and finally the point-like nuclear source with a magnitude
$m_V = 16.63$~\magsec and a half light radius of $r_h \sim 0\farcs054$,
values consistent with the ones given by Kormendy \etal (1996).
Note that it is yet impossible to disentangle between a flattening
of the surface brigtness profile of this nucleus inside $0.1\arcsec$
and a still increasing power law (Fig.~\ref{fig:showcusp}).
\end{itemize}
In the following modeling we will always use a distance of $10$~Mpc
for NGC~3115: this value is intermediate between 11.0 recently given
by Elson (1997) and the previously used one (9.24~Mpc - e.g. KR92).

\section{Two integral Jeans models}
\label{sec:jeans}

In this Section we wish to constrain the mass to light ratio in the
central $40\arcsec$ of NGC~3115 by comparing the observed kinematics
with Jeans dynamical models. We assumed a constant $M/L_V$ for
the luminous mass of the galaxy, and added any required additional
dark mass. We computed the second order non-centred projected velocity
moment $\mu_2 \equiv \sqrt{\tilde{V}^2 + \tilde{\sigma}^2}$
($\tilde{V}$ and $\tilde{\sigma}$ are the projected mean velocity and
velocity dispersion respectively) by solving the Jeans equations for a
two-integral distribution function $f(E, J)$. This is particularly
straightforward in the context of the MGE formalism as $\mu_2$ can be
evaluated through a single quadrature (see Emsellem \etal 1994).
Moreover, in the frame of a $f(E, J)$ dynamical model, $\mu_2$ does
not depend on the dispersion tensor anisotropy.  Observed $\mu_2$ can
only be determined with reasonable accuracy for data including higher
order moments.

\subsection{The inclination angle}

Taking into account the axis ratios of the disc components, the
inclination angle $i$ of the galaxy is restricted to a small range
between about $83\degr$ and $90\degr$ (edge-on). Note that the lower
limit is slightly model dependent as the MGE model cannot exactly
reproduce the thickness of the outer disc which is constant or
slightly rising outwards (see Capaccioli et al. 1988). We have built
Jeans models assuming different values for $i$, sampling the
interval between $83\degr$ and $90\degr$.  Varying the value of $i$
mostly affects the major-axis kinematics, where the flattened
components dominate the light distribution. The minor-axis $\mu_2$
profile remains nearly unchanged.  We can thus use the observed ratio
between the value of $\mu_2$ along the minor-axis and along the
major-axis as an indicator for the inclination angle, as it does not
depend on the mass to light ratio and anisotropy. A value of $i = 86\degr$
gives the best fit, and will be used for all following models.

\subsection{The stellar mass to light ratio}

Fig.~\ref{fig:ml} clearly shows that the central second order moment
is underestimated in the model with constant $M/L_V$.  In the
restrictive frame of our axisymmetric Jeans models with constant $M/L$
that we consider in this section, we could not find a reasonable fit to
the data in the central $3''$. Outwards from the central $3\arcsec$, we
obtained a rather good agreement with the $\mu_2$ profiles along both the
minor and major axes using $M/L_V = 6.5$ (Fig.~\ref{fig:ml}). 
\figstart{figure=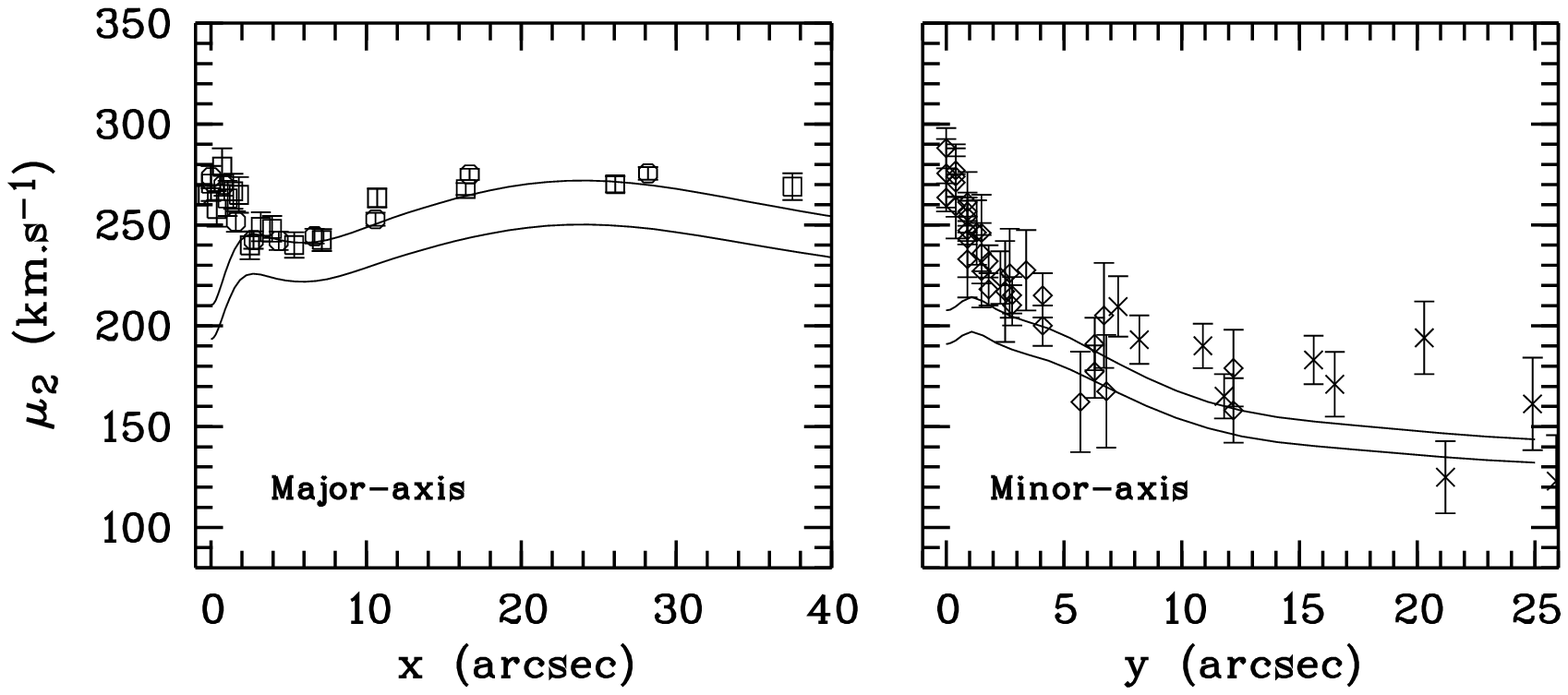,width=\hssize}
\caption[]{Major and minor axes second order moment profile (squares: vdM+94;
circles: F+96; losanges: KR92; crosses: IS82) compared with
the MGE jeans model with $M/L_V = 6.5$ (solid lines) and $M/L_V = 5.8$.}
\label{fig:ml}
\figend

\subsection{An estimation of the central dark mass}

As emphasized already in the previous Section and by previous studies
(see K+96 and references therein), the central dispersion gradient
calls for an increase of the mass to light ratio in the central few
arcseconds of NGC~3115. The FOS kinematics presented by K+96 then
implies that this takes place inside a radius of $\sim 0\farcs2$.
According to our Jeans models, the observed gradients in the
kinematics would lead to $M/L_V$ greater than 60
for the nucleus.  This strongly suggests the presence of a central
dark mass, which we denote by $\MBH$, that is larger than
$10^8$~\Msun\ (the mass of the nucleus with $M/L_V = 6.5$ would be $\sim
6.3 \times 10^7$~\Msun). In the following paragraphs we now provide some
constraints on the central mass concentration in the frame of a
$f(E,J)$ dynamical model.

The addition of a central point-like mass in the MGE model is
straightforward and the formalism is described in Appendix A of
Emsellem \etal (1994).  We built a number of models with masses
$\MBH$ ranging from $10^8$ to $2 \times 10^9$~\Msun\ and for
different $M/L_V$. In the following, we give the values of $\MBH$
which best fit each individual data set, as well as an indicative
range in an attempt to take into account the formal error bars of the
measured kinematics.  Note that fitting the central kinematics
requires to vary both $\MBH$ and $M/L_V$.  Fig.~\ref{fig:mbh}
presents the comparison between these models and data inside
$7\arcsec$ along the major-axis. 

Estimations of the true moments are sensitive to errors in the measured
high order Gauss-Hermite moments (e.g. $h_3$ and $h_4$, see Bender
\etal 1994): this may be the cause of the low central dispersion value
in vdM+94's data. If we ignore the two central points in the fit 
we find $\MBH \sim 6 \pm 1 \times 10^8$~\Msun\ and $M/L_V = 6.1 \pm 0.4$. 
The overall best fit to the vdM+94 data is obtained with $\MBH \sim
4 \pm 1.5 \times 10^8$~\Msun\ and $M/L_V = 6.2 \pm 0.3$. 
The TIGER data indicates a slightly higher value for the central dark mass of $\MBH
= 6.6 \pm 1 \times 10^8$~\Msun\ with $M/L_V = 6.5 \pm 0.5$.  Similar
comparisons have been done with the data of Fisher (1997) and
Bender \etal (1994) which both give similar values of $\MBH \sim
7.0 \pm 3 \times 10^8$~\Msun\ and $M/L_V = 6.8 \pm 0.5$.
\figstart{figure=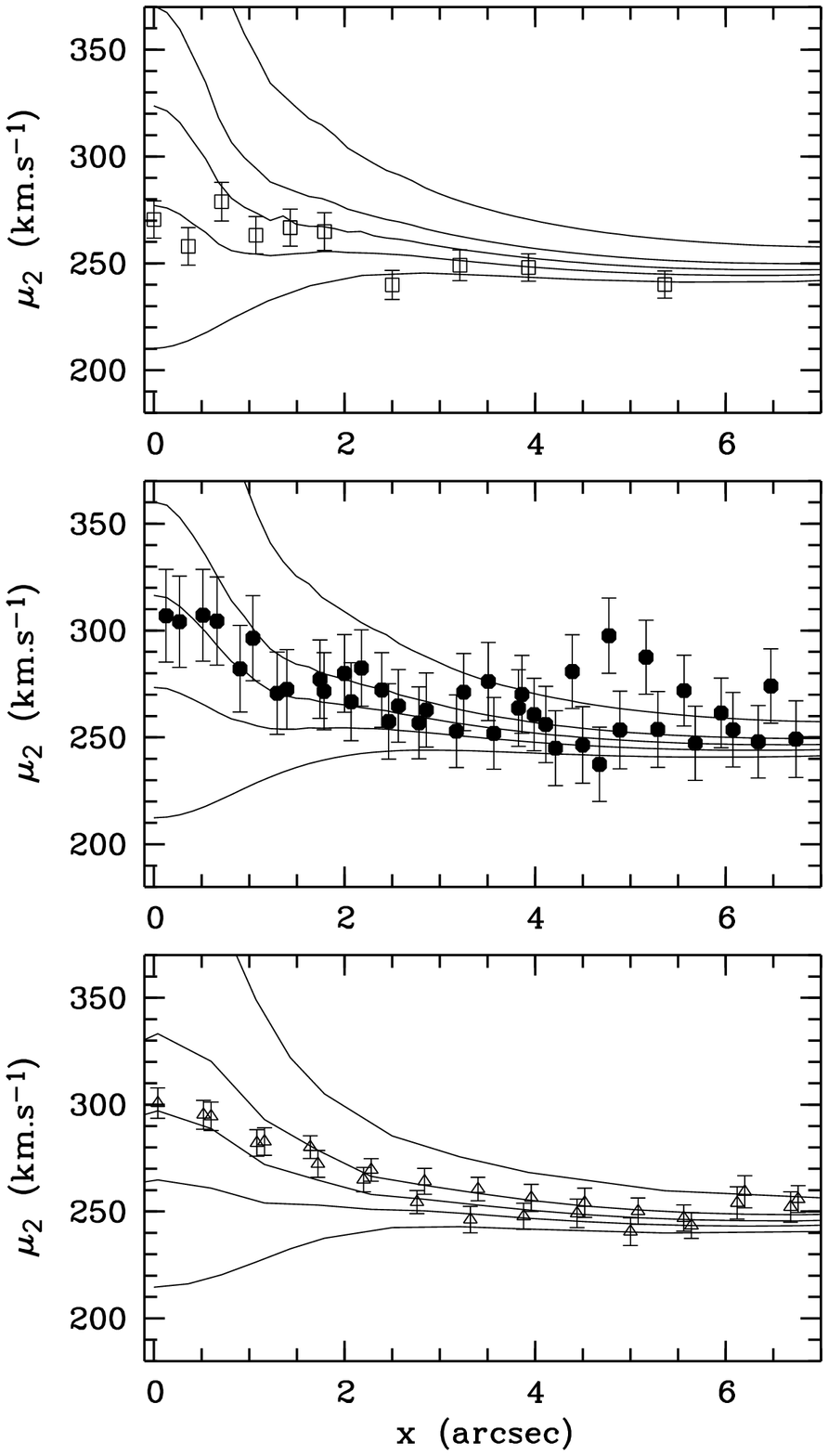,width=8.8cm}
\caption[]{Observed kinematics for 3 data sets (upper panel:  vdM+94,
middle panel: TIGER, lower panel: B+94) along the major-axis compared with
models with different black hole masses $\MBH$: 0, $3.7 \times
10^8$, $6.5 \times 10^8$, $1.0 \times 10^9$ and $2.0 \times
10^9$~\Msun.}
\label{fig:mbh}
\figend

For the global set of kinematical data, the best estimation for the
black hole mass is then $\MBH = 6.5 \pm 3.5 \times 10^8$~\Msun\ with
$M/L_V = 6.5 \pm 0.7$. Values as high as $2 \times 10^9$ quoted 
by K+96 are excluded at more than the 3$\sigma$ level
for the simple $f(E,J)$ case\footnote{The distance used in this paper for NGC~3115
is slightly higher than the one of K+96, which increases this
discrepancy.}.

\subsection{Dark matter at large radii}
 
At radii larger than $40\arcsec$ along the major-axis the observed
velocity and dispersion profiles are almost flat out to the last
measured point with $\langle V\rangle \sim 250$~\kms and
$\langle\sigma\rangle \sim 100$~\kms (Capaccioli \etal\ 1993).  The
Jeans model with $M/L_V = 6.5$ predicts a dispersion profile that is
consistent with the observed one at large radii, but underestimates
the velocity profile for $R > 70\arcsec$ by as much as 60\% at
$R=120''$.  As will be seen in Section~\ref{sec:QP}, this implies a
strong increase of the mass to light ratio at large radii.

\section{Two integral distribution functions}
\label{sec:2I}

\subsection{The method}
\label{sec:method}

In Sect.~\ref{sec:jeans}, we have restricted the range of values for
the central dark mass using Jeans models and estimations of the true
moments, and found a best fit with $\MBH = 6.5 \pm 3.5 \times
10^8$~\Msun. We wish now to compare the models directly to the
measured values, namely $V$, $\sigma$ and the higher order
Gauss-Hermite moments. This requires the knowledge of the full
distribution function. We therefore applied the method of Hunter \&
Qian (1993) to derive the even part of the distribution function $f(E,
J)$ corresponding to the axisymmetric MGE mass model of NGC 3115.

The even part $f_e$ of $f(E, J)$ was calculated with a
grid of $49 \times 38$ points in $(E,J)$-space designed to properly sample
the gradients of $f$ (particularly at $J(E) / J_{max}(E) \sim 1$
due to the presence of the discs).  The odd part
$f_o$ remains as a free function in our models with the constraint
that $f(E, J) = f_e(E, J) + f_o(E, J)$ is positive everywhere.
We have separated the contributions of the discs and the bulge in the
computation of the distribution function: this is achieved by using
the mass density of each component but including the total potential.

In order to derive the odd part $f_o$ we used the parametrization
introduced by van der Marel \etal (1994) following the work of
Dejonghe (1986,1987): we define a function $h_a(\eta \equiv J /
J_{max}(E))$ such as
\begin{equation}
h_a(\eta) = \left\{ \begin{array}{lll} 
& \tanh{\left(a \eta / 2\right)} / \tanh{\left(a / 2\right)} \;\;\; & (a > 0)\\
& \eta  & (a = 0)\\
& \left(2 / a\right) \mbox{arctanh}{\left( \eta \tanh{\left(a / 2\right)}\right)} \;\;\; & (a < 0)
\end{array}
\right.
\end{equation}
so that $f_o(E, J) = h_a(\eta) f_e (E, J)$.
In practice, we used different values of $a$ for the 
disc and bulge components. This was essential to fit 
the detailed observed kinematics.

The LOSVDs were derived from the obtained numerical distribution
function on a fine grid of more than 1000 points to allow a proper
sampling of the sky plane in the central $45\arcsec$. The areas of the 
projected LOSVDs (normalized by the projected luminosity
density) were always equal to 1 within $1\%$: this is
an {\em a posteriori} check of the validity of the overall
computations.  This grid of LOSVDs was then used to compute (via an
interpolation) the corresponding LOSVDs for each data set including
the effects of seeing convolution, pixel integration and spectral
resolution.  Finally the resulting LOSVDs were parametrized using
the Gauss-Hermite moments as well as the true velocity moments.
\figstart{figure=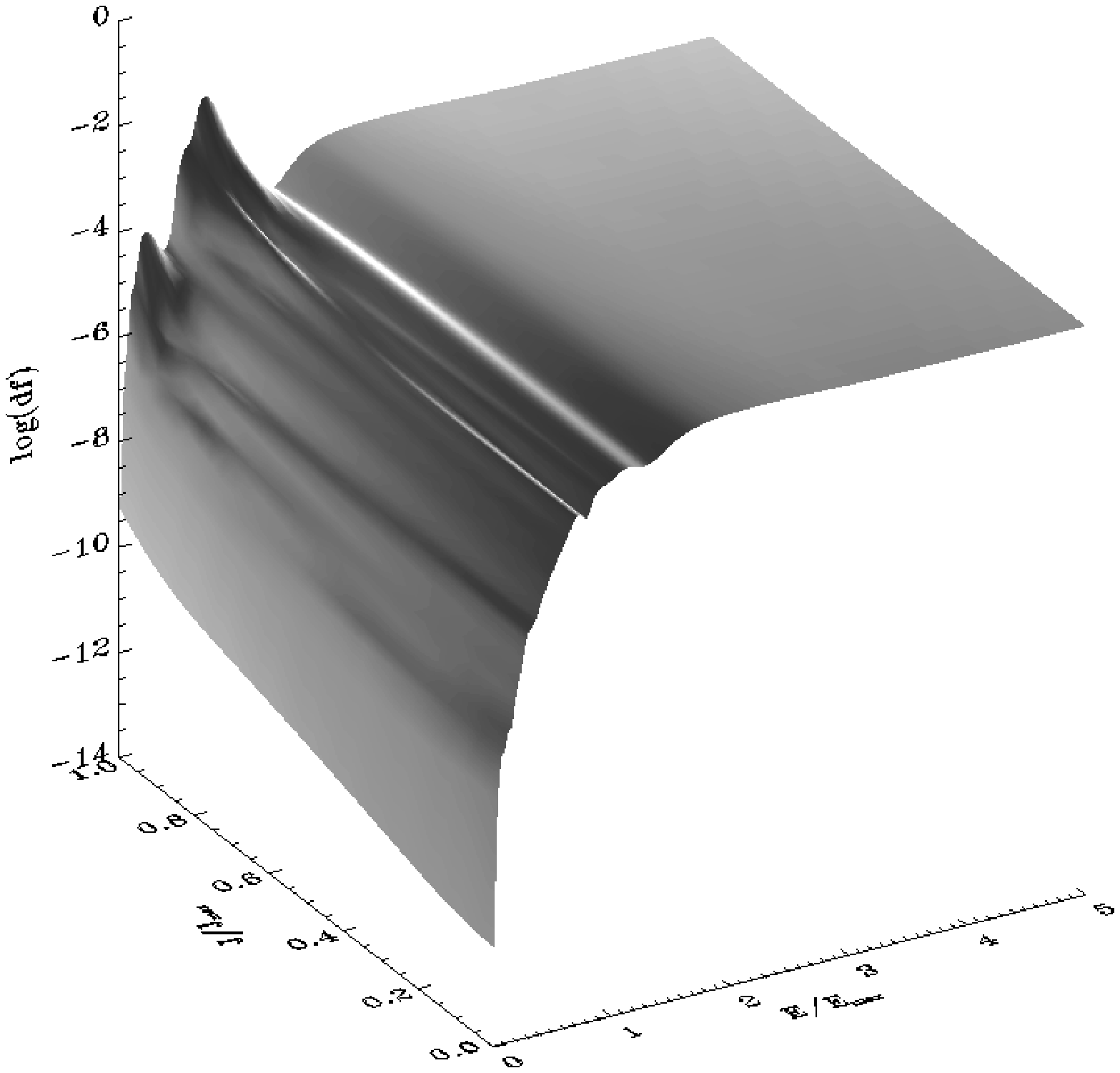,width=\hssize}
\caption[]{Even part of the distribution function obtained with
a mass model of NGC~3115 including a central dark mass of $\MBH = 1.0 \; 10^9$~\Msun:
the logarithm of the distribution function is plotted as a function
of the normalized energy $E / E_{max}$ and angular momentum
$J / J_{max}$. $E_{max}$ corresponds to the central potential
of the model when the central dark mass is excluded.}
\label{fig:df}
\figend

A number of models have thus been built, including different cusp
slopes and a range of values for the mass to light ratio $M/L_V$ and
for the central dark mass $\MBH$. In Fig.~\ref{fig:df}, we present
one example of such a distribution function obtained for
$\rho \propto r^{-1.5}$ and $\MBH = 1.0\times 10^9$~\Msun as a function 
of the normalized energy ($E_{max}$ corresponds to the central
potential of the model excluding the central dark mass) and angular
momentum ($J = J_{max}$ corresponds to circular orbits 
at a given energy). The double disc structure is clearly
visible as two bumps near the line of maximum angular momentum,
and the plateau at high energy corresponds to the cusp
(for a cusp slope $\gamma = -1.5$ and a keplerian potential
the distribution function is constant with energy, see Qian et al. 1995). 

\subsection{The inner 45 arcseconds}
\label{sec:45sec}
\figstart{figure=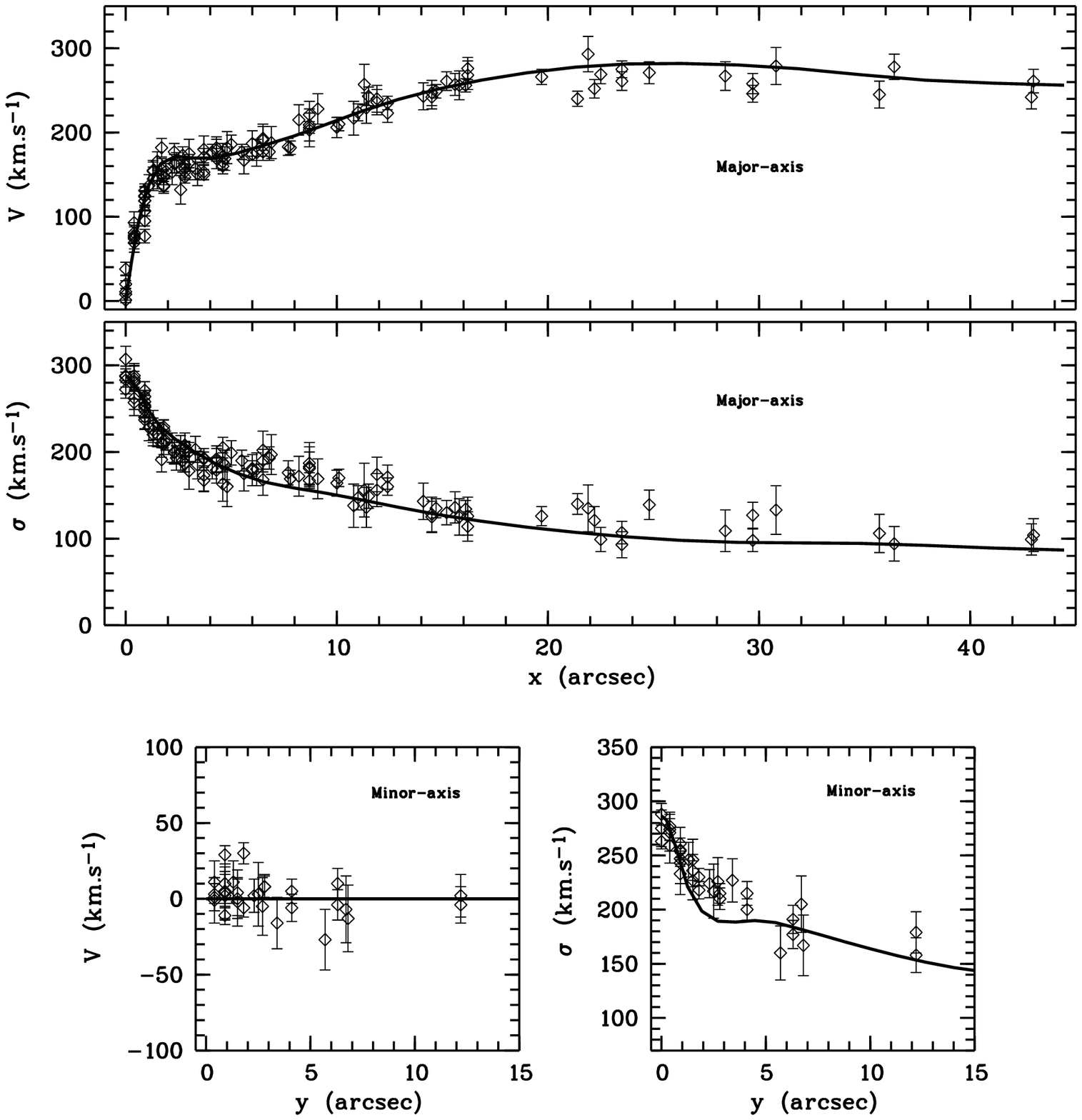,width=\hssize}
\caption[]{Observed kinematics (losanges) along the major-axis 
(two top panels) and the minor-axis (bottom panels) for the data of
KR92 and the best fit model with $\MBH = 0.94 \times 10^9$~\Msun.}
\label{fig:kor2I}
\figend

As shown in Sect.~\ref{sec:jeans}, an overall good fit to the data
inside $40\arcsec$ requires a rather high mass to light ratio $M/L_V
\sim 6.5$. For the best sets of data (namely vdM+94, CJ93, and KR92)
which cover these radii, we built models with different
central dark mass concentrations. In Fig.~\ref{fig:kor2I}
we present the KR92 data set which extends further than 
$40\arcsec$ and overimposed the best
fit model which includes a central dark mass of $\MBH = 0.94 \times
10^9$~\Msun and has a mass to light ratio $M/L_V = 6.1$. 
The major-axis velocity and dispersion (best gaussian $V$
and $\sigma$) profiles are well fit by this model even in
the inner part. The fit is also good along the minor-axis besides a significantly
lower dispersion around $3\arcsec$ where the central dark mass starts
to dominate the observed kinematics. The fit to the data of CJ93, which has a slightly
higher spatial resolution, is also excellent.

\figstart{figure=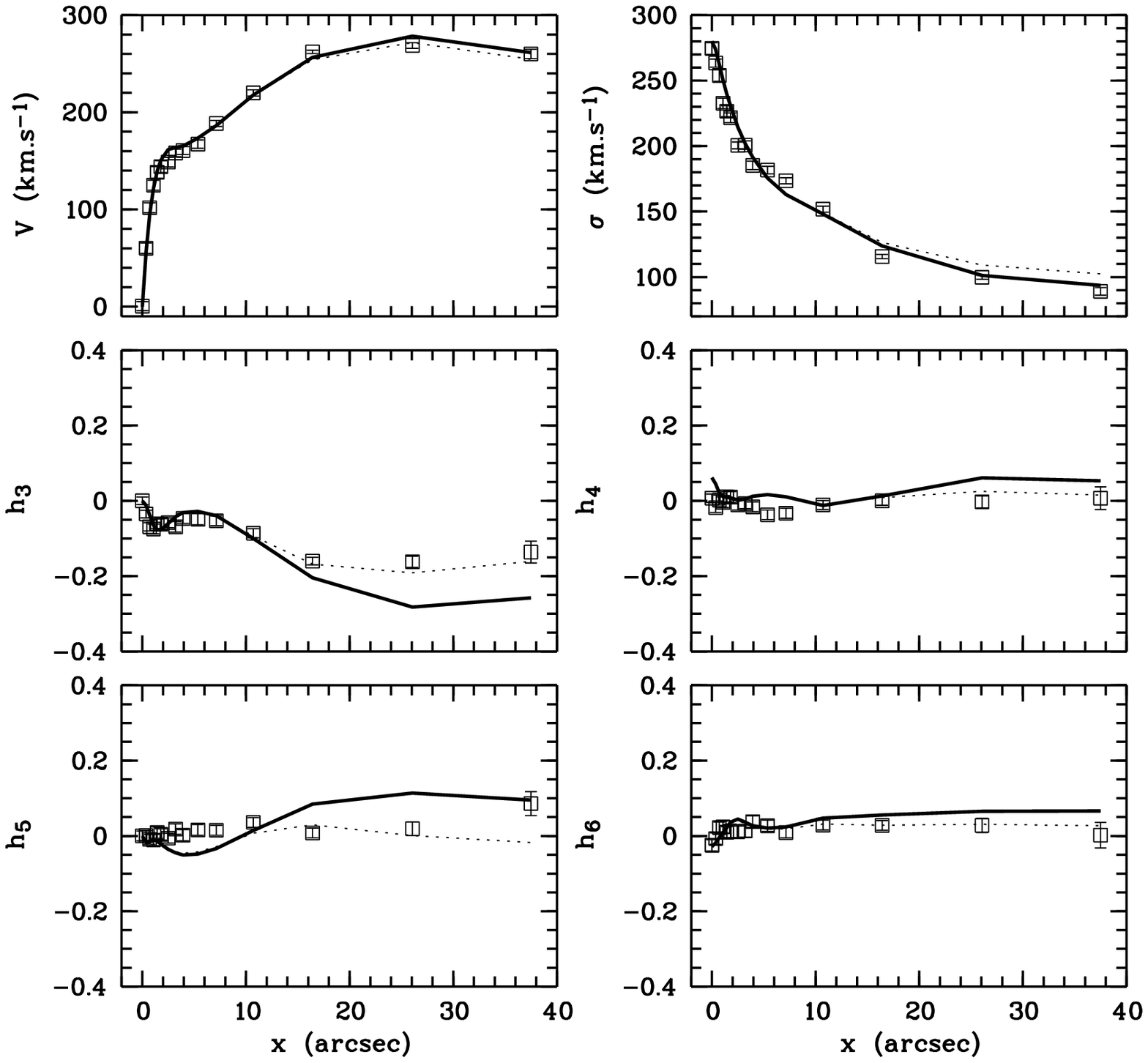,width=8.8cm}
\caption[]{Observed kinematics (open squares) along the major-axis 
for the data published in vdM+94, derived using the method described
by Rix \& White (1993). These data have been fit simultaneously on
both sides of the centre (see vdM+94 for details).  The best fit 2I
model with $\MBH = 0.94 \times 10^9$~\Msun\ is superimposed with
(dotted line) and without (solid line) spectral filtering.}
\label{fig:vdm2I}
\figend
In Figure \ref{fig:vdm2I} we present the fit to the vdM+94 data set.
In this case, the higher order Gauss-Hermite moments
could also be included. We could not obtain a reasonable fit to the $h_3$
parameter for which the 2I model predicts values around -0.25 at $R
\sim 25\arcsec$ while the observed one is $\sim -0.16$. It is possible
to reduce (in absolute value) the predicted $h_3$ by changing the odd
part of the distribution function, but at the expense of a
significantly lower velocity $V$. Note that the predicted $h_3$ does not depend on
the mass to light ratio (assuming it is constant with radius).

Uncertainties due to errors in the positioning of the slit or
to spectral filtering (see Sect.~\ref{sec:error}) of the LOSVDs
are far too small to fully account for this discrepancy.
The effect of the latter is shown in
Fig.~\ref{fig:vdm2I} where the predicted LOSVDs have been slightly
Fourier filtered before measuring the Gauss-Hermite moments: the fit
is then significantly better for all $h_3$, $h_4$, $h_5$, $h_6$
profiles, without changing $V$ and $\sigma$ significantly.  However,
the filtering required to explain the data is large considering the
signal to noise ($S/N \sim 20$ per 10 km.s$^{-1}$) and spectral
resolution ($\sigma_{inst} \sim 23$~km.s$^{-1}$) quoted by vdM+94.
We then performed some extensive simulations to derive the effect of the deconvolution
method (Fourier fitting of the LOSVD) on $V$ and $h_3$: it is indeed significant 
for $R \sim 25\arcsec$, where it can reach $15$~\kms and $\sim 0.04$ respectively.
The data of vdM+94 are thus consistent with
$h_3 = -0.2$ at $R \sim 25\arcsec$, still too low in absolute value
to be reconciled with our two-integral model.
There are finally uncertainties associated with
the data as illustrated in Fig.~\ref{fig:newvdm2I}: the same 2I model 
is compared with the kinematics obtained by vdM+94 but
this time derived using a one side Fourier-fitting technique (see
vdM+94 for details). The fits to the higher order moments are good,
although the predicted $V / \sigma$ is now too low.  

\figstart{figure=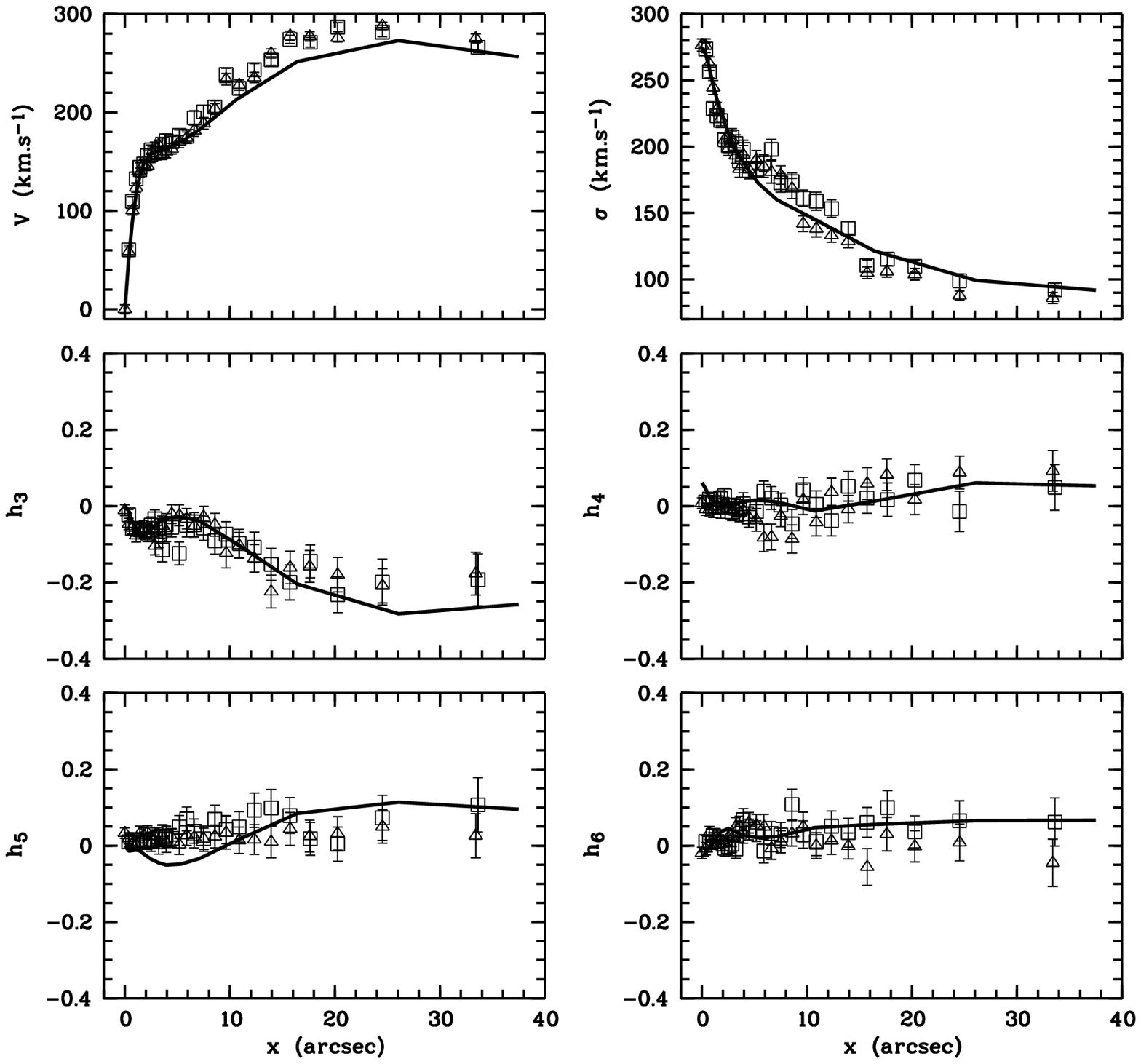,width=8.8cm}
\caption[]{Observed kinematics (open squares) along the major-axis 
for the data of vdM+94, derived from a Fourier fitting method. Here the
data has been derived on each side of the centre independently
(north-east side: open triangles; south-west side: open squares).
The best fit 2I model with $\MBH = 0.94 \times 10^9$~\Msun\
is superimposed (solid lines).}
\label{fig:newvdm2I}
\figend

\subsection{The central arcseconds}
\figstart{figure=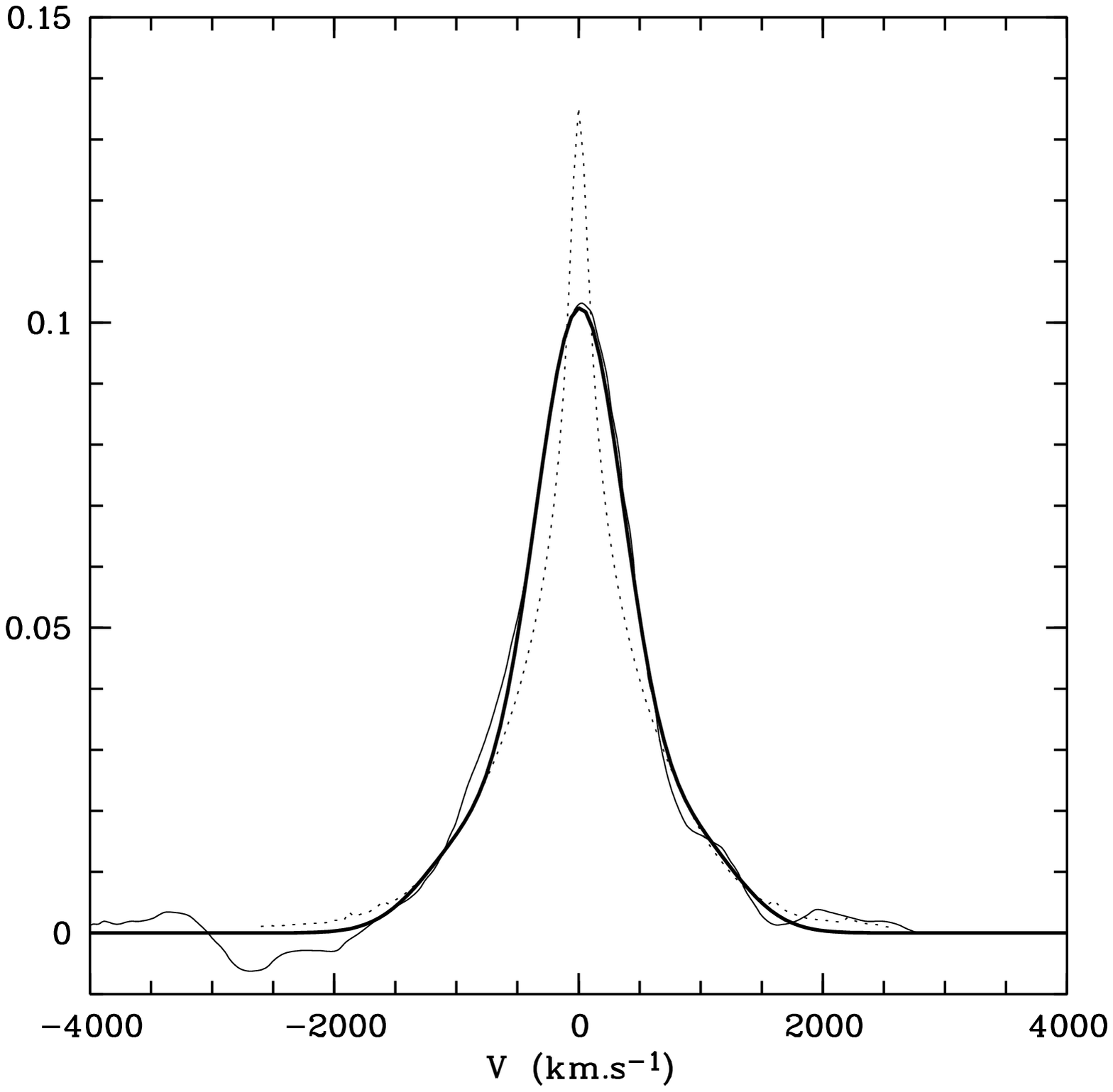,width=8.8cm}
\caption[]{Central LOSVDs: observed FOS (thin solid line, shifted to zero 
velocity), directly predicted from model M1 (dotted line), and the
same obtained via FCQ and using the characteristics of the FOS central
spectrum (thick solid line).}
\label{fig:hstlos}
\figend

The observed kinematics in the central arcseconds is very sensitive to
the spatial resolution, particularly in the case of NGC~3115 for which
both the photometry and kinematics exhibit large central gradients.
But as we have shown in Sect.~\ref{sec:h4}, the procedures used in the
analysis of the data do also affect the final result.
In the following paragraphs, we used the data provided by 
the FOS/HST and SIS spectrographs in parallel with
the (lower resolution but) two-dimensional TIGER kinematics to
constrain the free parameters of the dynamical models. We again
examined a large range of different models, varying the central dark
mass, the mass to light ratio, the central cusp slope as well as the
odd part of the two-integral distribution function. For reasons of
clarity, we only present here a few of these models which fit the data
well.

\figstart{figure=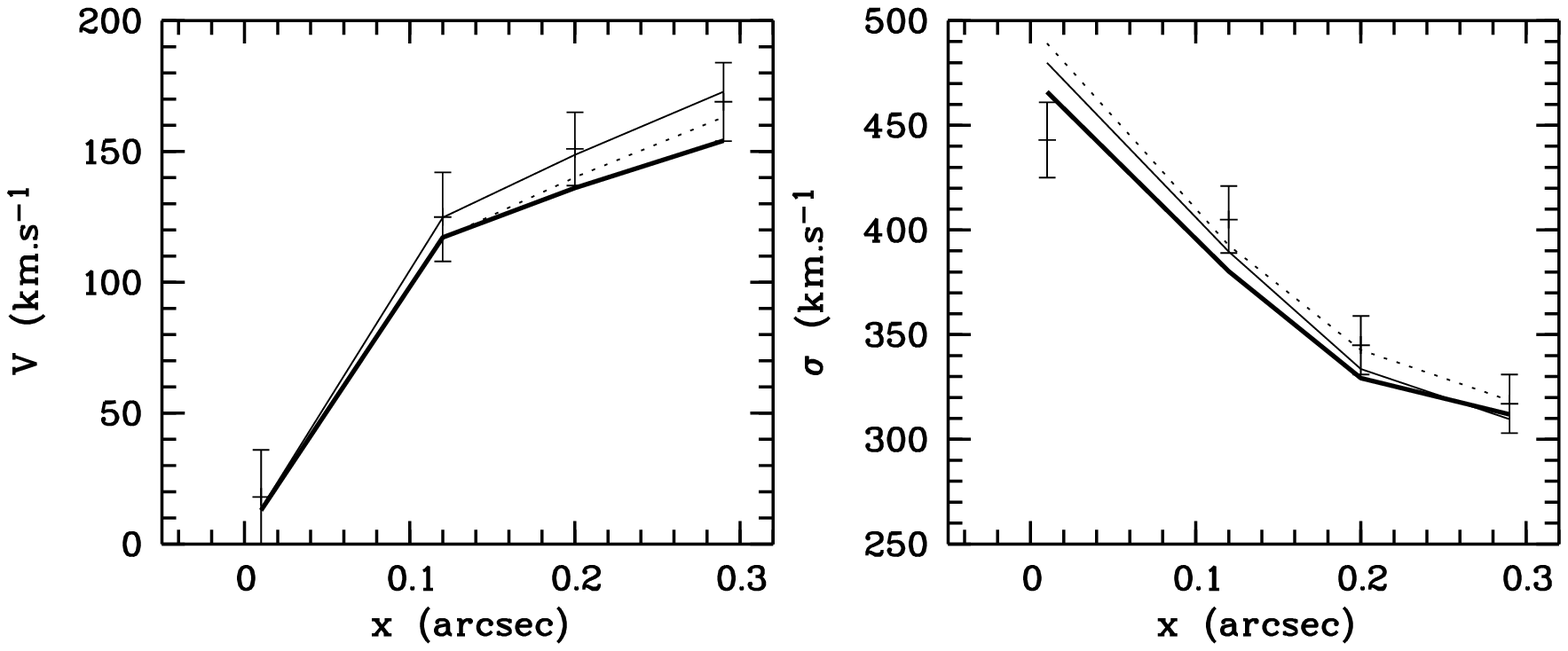,width=8.8cm}
\caption[]{Observed FOS kinematics along the major-axis (K+96, crosses),
compared with the velocity and dispersion profiles predicted by models
M1, M2 and M3 (thin solid, dotted and thick solid lines respectively)
via FCQ and using the characteristics of the FOS data.}
\label{fig:hst}
\figend

\subsubsection{Best fit models}
\figstart{figure=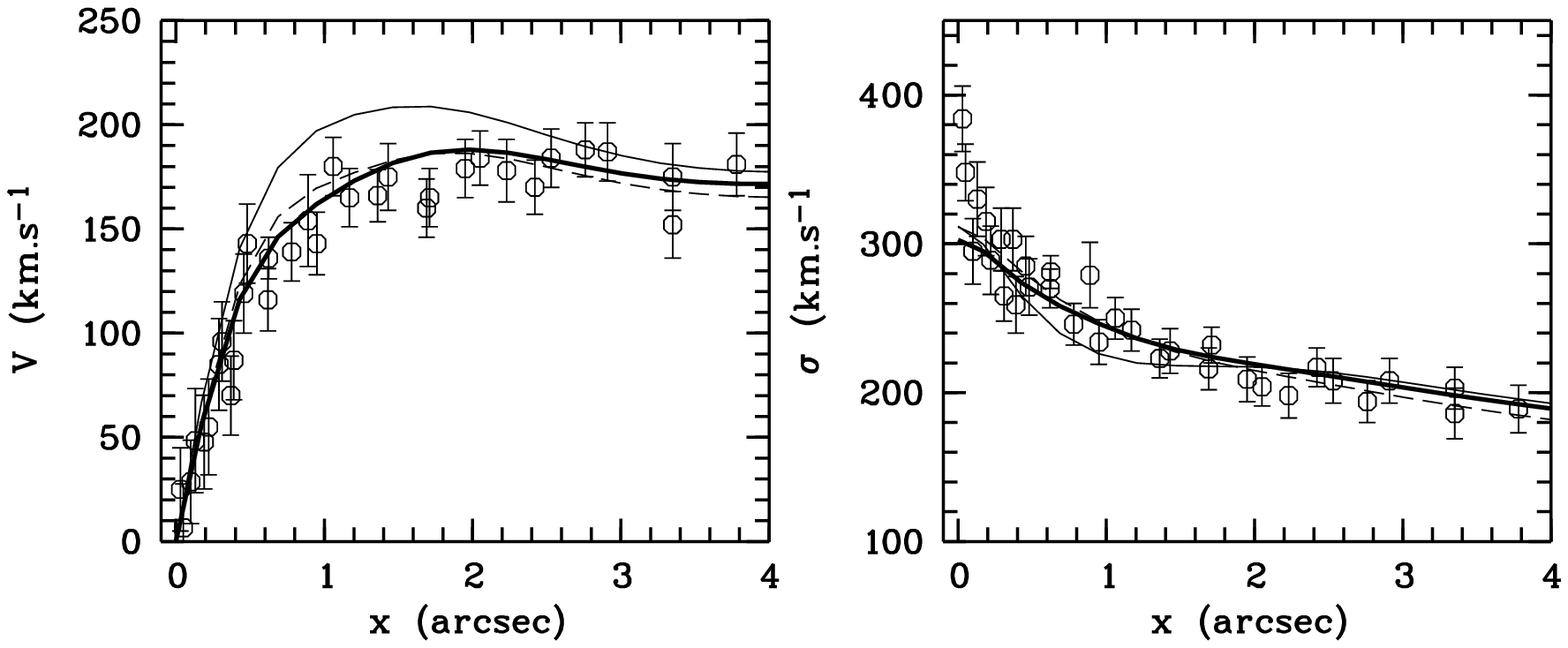,width=8.8cm}
\caption[]{Observed SIS kinematics along the major-axis (K+96, circles), with
the thin solid, dashed and thick solid lines corresponding to
predictions of models M1, M2 and M3 respectively.}
\label{fig:sis}
\figend

All our two integral models predict rather high $h_4$ values ($>
0.08$) when $\MBH$ is a few $10^8$~\Msun.  Our best fit model to the
FOS data only has a cusp with $\rho \propto r^{-1.5}$ and a
flattening of 0.7, a central mass $\MBH = 1.0\times10^9$~\Msun\ and
$M/L_V = 6.5$ (model M1, see Fig.~\ref{fig:hst}), although the cusp
slope and flattening are not well constrained by the present data.
In Fig.~\ref{fig:hst}, we present the kinematical profiles
obtained from the FOS spectra and model M1. The overall agreement is
good, but our two integral model predicts a slightly higher
central velocity dispersion with $\sigma(0) = 480$~km.s.$^{-1}$.  We
should note however that the best gaussian fit to the LOSVD obtained
by K+96, and presented in their Fig.~3 has $\sigma(0) \sim
489$~km.s$^{-1}$, significantly larger than the one derived via FQ,
but consistent with our value. This point is emphasized in
Fig.~\ref{fig:hstlos} where we now present the comparison between the
FOS LOSVD obtained by K+96 and the one predicted by model M1.  The
agreement is excellent, considering that the wiggles present in the
FOS central LOSVD are almost certainly not real (as are the negative
points), but resulting from the presence of noise in the FOS spectrum
as well as the template mismatching. It is striking to see the
difference between the LOSVD directly predicted from model M1 and the
one retrieved via FCQ using the same spectral filtering than for the
FOS data: this forces us here to underestimate $h_4$ and
correspondingly to overestimate the gaussian dispersion (see
Sect.~\ref{sec:h4}).
 
Model M1 includes a rapidly rotating nuclear disc with $a = 5$ (see
Section~\ref{sec:method}).  This model predicts velocities which are
clearly too high at the SIS resolution, as shown in
Fig.~\ref{fig:sis}.  Kormendy et al. (1998) noticed the fact that, for
SIS data of NGC~3377, FCQ seems to provide slightly higher
velocities than FQ by about 10\% (e.g. see their Fig.~15). If this is
also the case for NGC~3115, it could reconcile our model M1 with the
SIS data.
\figstart{figure=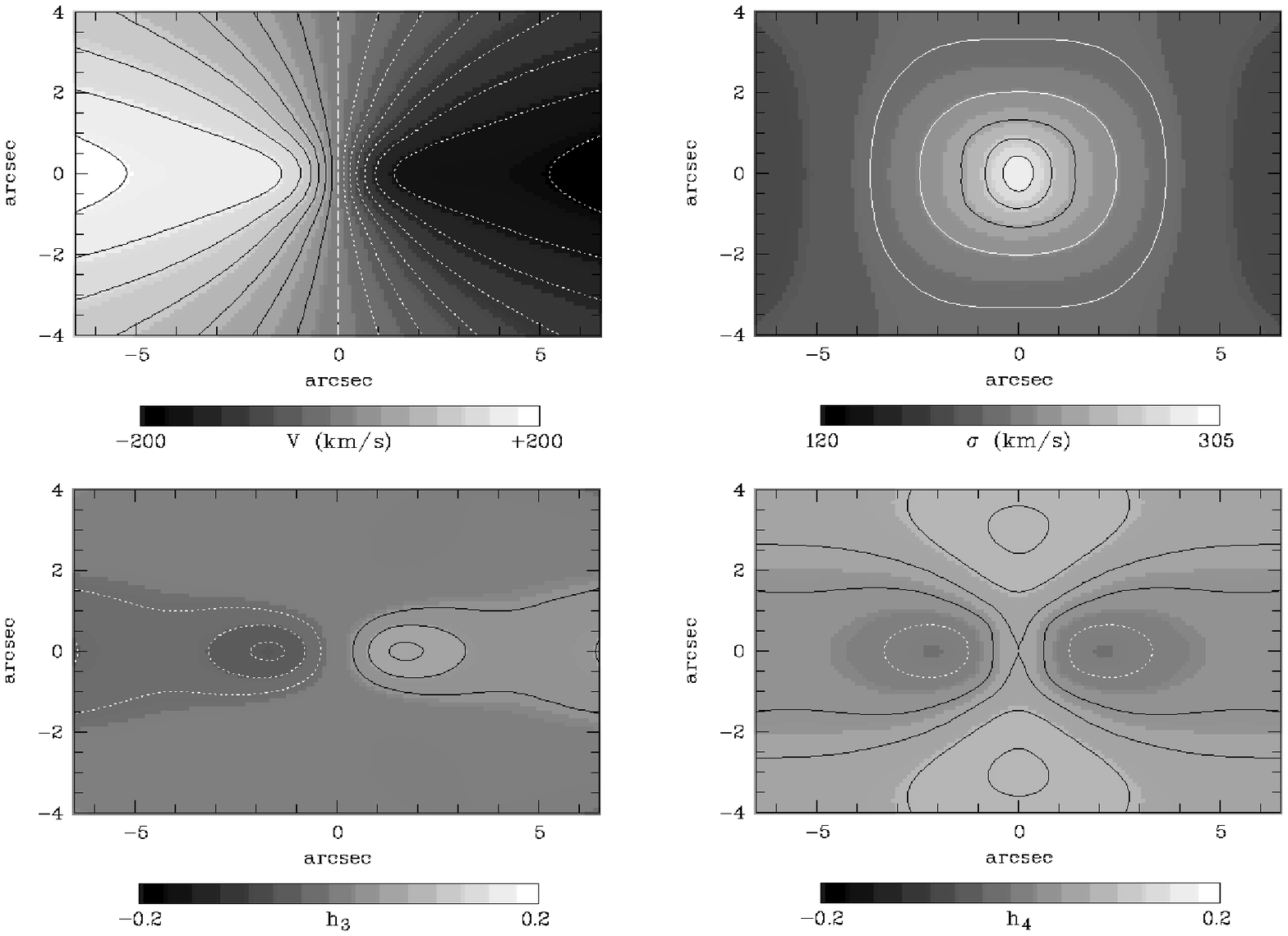,width=\hssize}
\caption[]{Predicted 2D kinematics for the TIGER setup from model M1. From
top to bottom, left to right: $V$, $\sigma$, $h_3$ and $h_4$. Contours
are the same than in Fig.~\ref{fig:tigmaps}.}
\label{fig:tiger}
\figend

We calculated other models where the parameter $a$ was
significantly increased (up to $a = 100$) for stars with high energy
(close to the centre as we deal with positive energies): in this
region the total distibution function tends towards the maximum
streaming two-integral model.  Fig.~\ref{fig:hst} and
Fig.~\ref{fig:sis} present two other models, respectively M2 which
corresponds to $\MBH = 1.24 \;10^9$~\Msun and $M/L_V = 5.4$, and M3
which has $\MBH = 0.94 \;10^9$~\Msun and $M/L_V = 6.1$. Although
model M2 is a better fit to the FOS velocity and dispersion profiles,
the higher black hole mass implies $h_4 = 0.134$, a value at the upper
limit of the error bar mentioned by K+96. Model M3 is the best fit to
the SIS data, and represents the best compromise to the HST data with
a measured central $h_4 = 0.094$. These models also show that the detailed 
kinematical structure inside $0\farcs3$ remains ill constrained. Although
stellar orbits may be biased towards near maximum angular momentum values
in the central 10~pc, it is thus yet improper to discuss the internal kinematical structure 
of the point-like nucleus detected in the photometry (Sec.~\ref{sec:MGE}), all the more
since its luminosity distribution is uncertain as well 
(see Fig.~\ref{fig:showcusp}).

\subsubsection{The TIGER data}

We finally compare the three best fits (M1, M2 and M3) to the
kinematics obtained with TIGER. Although the resolution of these data
is significantly lower than the SIS data, it still represents a
good constraint as it homogeneously covers a two-dimensional field on
the sky. As seen in Fig.~\ref{fig:tigmaps}, the TIGER isovelocities
are significantly flattened.  This suggests again that the nuclear
disc is rapidly rotating.  The contrast between the disc and the
central spheroid is however weaker than in the case of the Sombrero
galaxy for which very asymmetric LOSVDs (and therefore high $h_3$
values) are observed at the centre (Wagner et al. 1989; Emsellem et
al. 1996). All three models, M1, M2 and M3, fit reasonably well the
TIGER kinematical maps, with M1 giving the smallest residuals.
The kinematical maps corresponding to model M1 are presented in 
Fig.~\ref{fig:tiger}.

The results presented in the previous Sections, using two integral $f(E,J)$ 
models to fit the observed kinematics, can thus be summarized in a few points:
\begin{itemize}
\item There is a significant discrepancy between the observed
and predicted $h_3$ values at a radius of about $25\arcsec$.
\item The fit of the central FOS kinematics requires the nuclear disc to 
have a nearly maximum streaming distribution function inside $1\arcsec$.
\item All the available kinematics in the central arcseconds 
can be rather well fit by a two-integral model with constant mass
to light ratio.  The best constrained parameters are the central dark
mass and the mass to light ratio: considering the error bars in the
different data sets, our best estimations are $\MBH =
0.94^{+0.35}_{-0.2} \;10^9$~\Msun and $M/L_V = 6.1^{+0.5}_{-0.7}$.
These values can be refined using the higher order Gauss Hermite
moments in the central arcsecond. However the observed
kinematics are very sensitive to the details of the instrumental setup
and the analysis procedures.
\end{itemize}

We could in principle reconcile our 2I model with the observed kinematics
along the major-axis. The input mass model has been indeed derived from a 
$V$ band image which may not reflect the true mass 
distribution. We have for example assumed that
the bulge and disc components have the same mass to light ratio. This
is certainly not true since they do not share the same colour profiles
(Fisher et al. 1996). A difference of 0.1 magnitude in $B - R_c$
as typically observed between the bulge and the disc regions in NGC~3115
(Fisher \etal 1996) corresponds to a maximum change of about 5\% 
in $M/L_V$ (Worthey \etal 1994), too small an effect to account for the 
discrepancy in $h_3$. In the following, we wish to examine 
another explanation: the distribution function depends on a third integral of motion. A two
integral distribution function $f(E, J)$ imposes that $\sigma_R =
\sigma_z$ everywhere.  This may not be the case for the S0 galaxy
NGC~3115, as for our own Galaxy. This is also supported by derivations
kindly made by Roeland van der Marel, using the formalism
described in de Bruijne et al. (1996) which allows a rather general 
geometry of the velocity ellipsoid. Although these models cannot 
mimic the complex morphology of NGC~3115, they clearly suggest that,
for very flattened components, low $h_3$ values of the order of $-0.2$ 
could be obtained by relaxing the constraint of a two-integral distribution function.

\section{A three integral distribution function}
\label{sec:QP}

In this Section, we will construct a three-integral model using a
quadratic programming technique.  
With the three integral modeling, we wish to approach two questions
\begin{enumerate}
\item Is it possible to fit the data without a central dark mass, with a distribution
function that has the appropriate three integral structure? In this context, we
recall the well-known case of M87, where Binney \& Mamon (1982)
succeeded in fitting the (then available) data with a distribution
function that was very anisotropic.
\item What is the phase-space structure of the best global 3I model?
\end{enumerate}
This is an opportunity to present the use of the Quadratic Programming
technique on a complex object, using a complete set of two-dimensional
kinematical data. It should be clear from the outset, however, that we 
chose to produce analytical distribution functions. These implies that our
third integral must be approximate. In that sense, the three integral
models we present here must be seen as belonging to the class of models
with approximate third integrals.

\subsection{The method}

The essence of the method relies on
the fact that most, if not all, observable kinematic quantities can be
expressed as linear functionals of the distribution function, if the
potential is given.  This means that if we approximate the
distribution function $f$ as a sum of some conveniently chosen basis
functions $f_i$:
\begin{equation}
f = \sum_i c_i f_i(I_1,I_2,I_3),
\label{dfqp}
\end{equation}
then the same relation holds between the (e.g.) projected mass
density $\rho_p$ and the projected mass density of the components
$\rho_{p,i}$
\begin{equation}
\rho_p = \sum_i c_i \rho_{p,i},
\end{equation}
or a similar relation for the projected pressures:
\begin{equation}
\rho_p(\sigma_p^2+\langle v_p\rangle^2) = \sum_i c_i 
[\rho_p(\sigma_p^2+\langle v_p\rangle^2)]_i.
\end{equation}
If we stick to the mass density $\rho$ for simplicity, 
we can construct a $\chi^2$ variable with the observed values
$\rho_l$ as follows:
\begin{equation}
\chi^2 = \sum_l w_l\left(\rho_l-\sum_ic_i\rho_{i,l}\right)^2,
\end{equation}
with $\rho_{i,l}$ the value of the mass density at the $l-$th point
for component $i$, and the $w_l$ associated weights. Clearly this $\chi^2$
is quadratic in the coefficients $c_i$.  Inclusion of other
observables is trivial. 

The minimisation of such a $\chi^2$ will produce a set of
coefficients, which will yield a distribution function (\ref{dfqp})
that will not necessarily be positive everywhere.  Therefore, one must
include constraints
\begin{equation}
\sum_i c_i F_i[(I_1,I_2,I_3)_m] \ge 0, \qquad m=1,\ldots,M,
\end{equation}
expressing the positivity of the distribution function on a grid in
phase space with points $(I_1,I_2,I_3)_m$. This formulates a problem
of quadratic programming, which is a well-known problem
for which efficient algorithms exist.

In the present case, we use the luminosity model as presented in section
\ref{sec:MGE}. As discussed in Sect.~\ref{sec:jeans}, a constant M/L model will not work
at large radii. This is indicated by the rotation curve,
which is almost flat up to $200''$ (Capaccioli et al. 1993). 
We therefore construct a model for
the dark mass that preserves the form of the luminous mass (say, for
simplicity of the argument, ellipses with fixed axis ratios and
semi-major axes $a$) and that has the radial dependence
\begin{equation}
\rho_{\rm tot}(a) = \rho_{\rm lum}(a)(1+ C a^p).
\label{dmassfunc}
\end{equation}
We used $p=1.5$, and the coefficient $C$ is such that
\begin{equation}
{\rho_{\rm tot}(a) \over \rho_{\rm lum}(a)} = 30, \qquad a=30 {\rm kpc}.
\end{equation}
This prescription produces a flat rotation curve.  In practice, we
adopt the same shape for the dark matter as for the luminous mass
density, by specifying the relation (\ref{dmassfunc}) for all the
radial functions in the harmonic expansion of $\rho_{\rm lum}$.
In Table~\ref{tab:ml}, we present the dependence of 
the mass to light ratio of the three-integral model 
on the radius in the equatorial plane. For the sake of comparison, we
will also present in Sect.~\ref{sec:res3I} a model where the mass to light ratio
was forced to stay constant (besides the central dark mass).
\begin{table}
\caption[]{Mass to light ratio of the three-integral model versus the radius in
arcsec or kpc. The contribution of the luminous mass is given in the right column}
\begin{center}
\begin{tabular}{|r|r|rr|}
\hline
$R$ & $R$ & $M/L_V$ & $M_{lum}$ \\
(kpc) & ($\arcsec$) & & (\%) \\
\hline
\hline
0.1  &   21  &  5.8 &   100 \\
1.5  &   30 &   6.1  &   90 \\
2.9  &   60 &    9.9  &   79 \\
4.4  &   90  &   13.9  &   68 \\
5.8  &   120  &   19.9  &   59 \\
7.3  &   150 &    26.0 &    52 \\
8.7  &  180  &   32.5  &   46 \\
10.2 &    210 &    39.5  &   41 \\
11.6  &   240  &   46.9  &   38 \\
13.1  &   270  &   54.4 &    35 \\ 
14.5  &  300   &  63.1  &   33 \\
\hline
\end{tabular}
\end{center}
\label{tab:ml}
\end{table}

The calculation of the potential is straightforward, when the harmonic
expansion of the total mass density is given. In practice, we
interpolate the potential on a grid. Since it is our goal to explore
what three integral dynamical model contributes to the modeling of
NGC~3115, we need an expression for the third integral $I_3$. This is
done by fitting the potential with a St\"ackel potential, and adopting
the third integral from the St\"ackel case as our approximation of the
third integral.  Thus, our models belong to the class of 3I models
that use an approximation for the third integral. The procedure is
outlined in more detail in Dejonghe \etal (1996).

Finally, we include a central singularity. It is a softened
$1/r$ singularity (hence a Plummer sphere), with a softening length
of the order of 50000 AU ($\sim 0.25$~pc or $\sim 5$ mas, about forty times
smaller than the FOS aperture). In the transition region where the
gravitational force of the black hole is of comparable strength as the
gravitational force of the core of the galaxy, there is no third
integral and the St\"ackel approximation of the potential is probably
invalid. However, as argued in Sect.~\ref{sec:jeans}, we adopt as
our working hypothesis that the spheroidal central component is
two-integral. Hence, it will suffice if we produce three-integral models
that do no penetrate too deep into the centre, where the model
thus remains two-integral. It is however important to note that since
orbits are non-local, the three-integral nature of our models
significantly extends the probed space of solutions.

The data we use to build the $\chi^2$ are photometric and kinematical.
As for the photometry, we produce a series of data points taken from a
(logarithmically spaced) grid of the deprojected luminous mass density
as obtained in Sect.~\ref{sec:MGE}. The kinematical data for this modeling
part are twofold.
\begin{itemize}
\item Outside $50\arcsec$ along the major-axis and $8\arcsec$
along the minor-axis, the effect of pixel integration and seeing convolution
is negligible, and we assumed that the gaussian velocity
and dispersion are good approximations for the two first true moments. 
We thus selected a set of points from the above mentioned sources.
It is impractical (yet) to include all data points, 
while some are, mildly inconsistent.
\item The rest of the galaxy (mainly the central part and
the major-axis inside $50\arcsec$) was sampled using
the estimations of the true moments given by the TIGER data 
presented in Sect.~\ref{sec:tigdata} and the published kinematics of vdM+94.
This of course required to include the instrumental setup in the $\chi^2$.
\end{itemize}

As for the components, we use two families. The first one
is an extention of Fricke components,
and has distribution functions of the form
\begin{equation}
F_{pmnE_0 J_{0}}^{\pm} = (E-E_0)^p (J-J_{0})^{2m} I_3^n
\end{equation}
wherever $\pm J\ge J_{0}$ and $E>E_0$. They are zero elsewhere.
The exponent $m$ and $n$ are an integer, but $p$ can be real.  In
general the parameter $p$ controls the central concentration of the
component, the parameter $q$ controls the amount of angular momentum,
while $n$ is indicative of the dependence on the third integral.  The
cut-off binding energy $E_0$ is needed to limit the extent of the
components, for further enhanced flexibility.  Also, since the
potential is almost singular at the centre, large values of $p$ (thus
components that are concentrated) produce impossibly high mass
densities at the centre.  Hence, we rather control the central
concentration with the parameter $E_0$, and adopt low or even
fractional values for $p$. Finally, the cut-off angular momentum
$J_{0}$ prohibits the orbits to penetrate too deep into the
potential well, where the third integral may not exist.

The second family is designed to produce very thin 2I discs in the
equatorial plane of the galaxy (Batsleer \&  Dejonghe 1995), and has
members of the form
\begin{equation}
F_{pqs}^{\pm} = (S_{z_0}(J))^p J^{2q} (E-S_{z_0}(J))^s
\end{equation}
wherever $\pm J\ge0, E>S_{z_0}(J)$. They are zero elsewhere.  The
parameters $p$, $q$ and $s$ can be real. The quantity $S_{z_0}(J)$
is the maximum binding energy that a star with angular momentum $J$
can have if its distance from the equatorial plane remains below
$z_0$.  If $z_0=0$, $S_0(J)$ is the binding energy of a circular
orbit with $J$. The parameters $p$ and $q$ have the same meaning as
in the first family, while $s$ controls the concentration of the
component towards the equatorial plane.

\subsection{Results}
\label{sec:res3I}

\figstart{figure=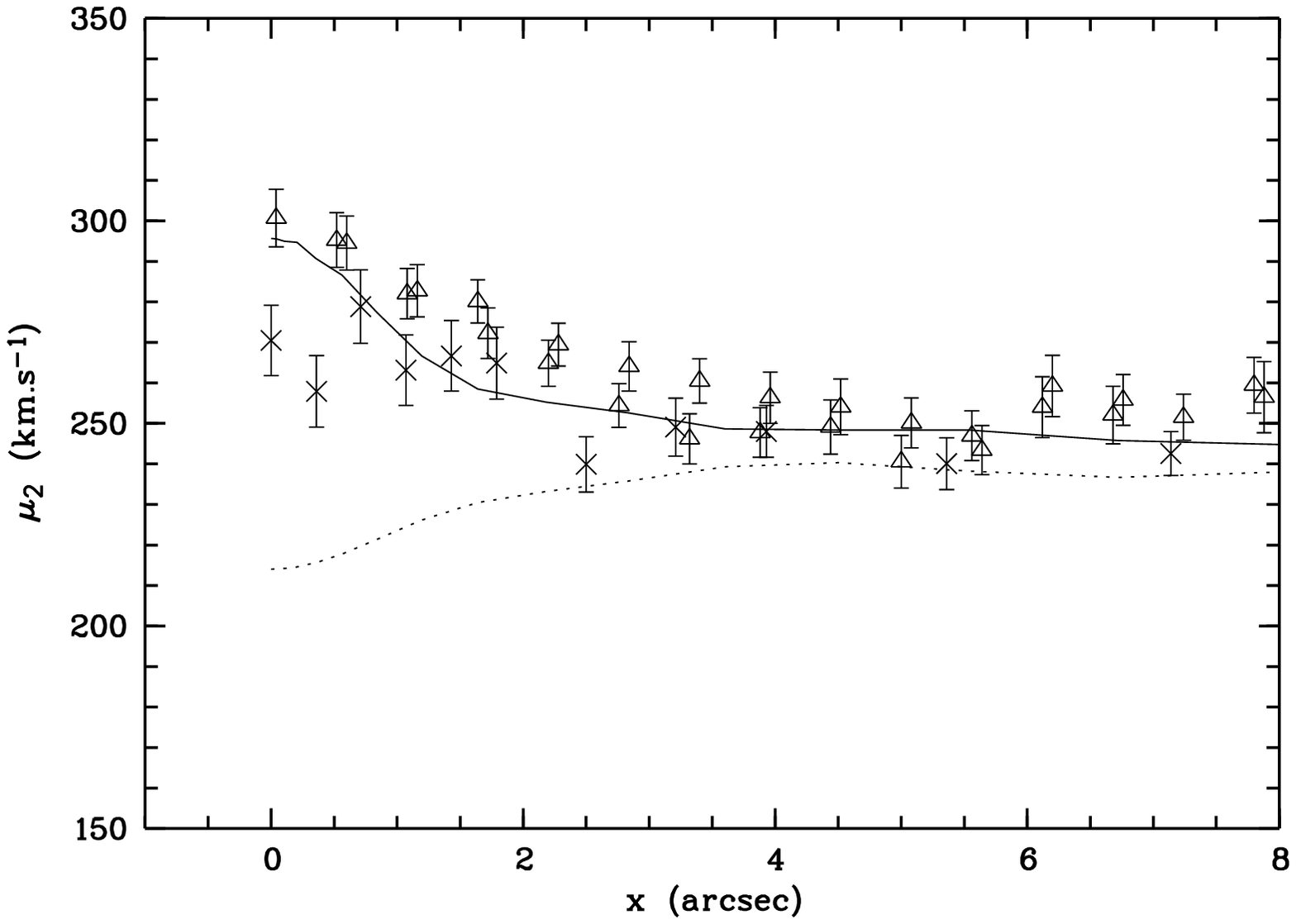,width=\hssize}
\caption[]{Data points from B+94 (triangles) and vdM+96 (crosses), and the 
best fit (3I models) with a $6.5\times10^8$~\Msun black hole (solid line) and best fit for
a model without black hole (dotted line). The models have been convolved to correspond to
B+94 data.}
\label{fig:compnobh-bh-centr}
\figend
\figstart{figure=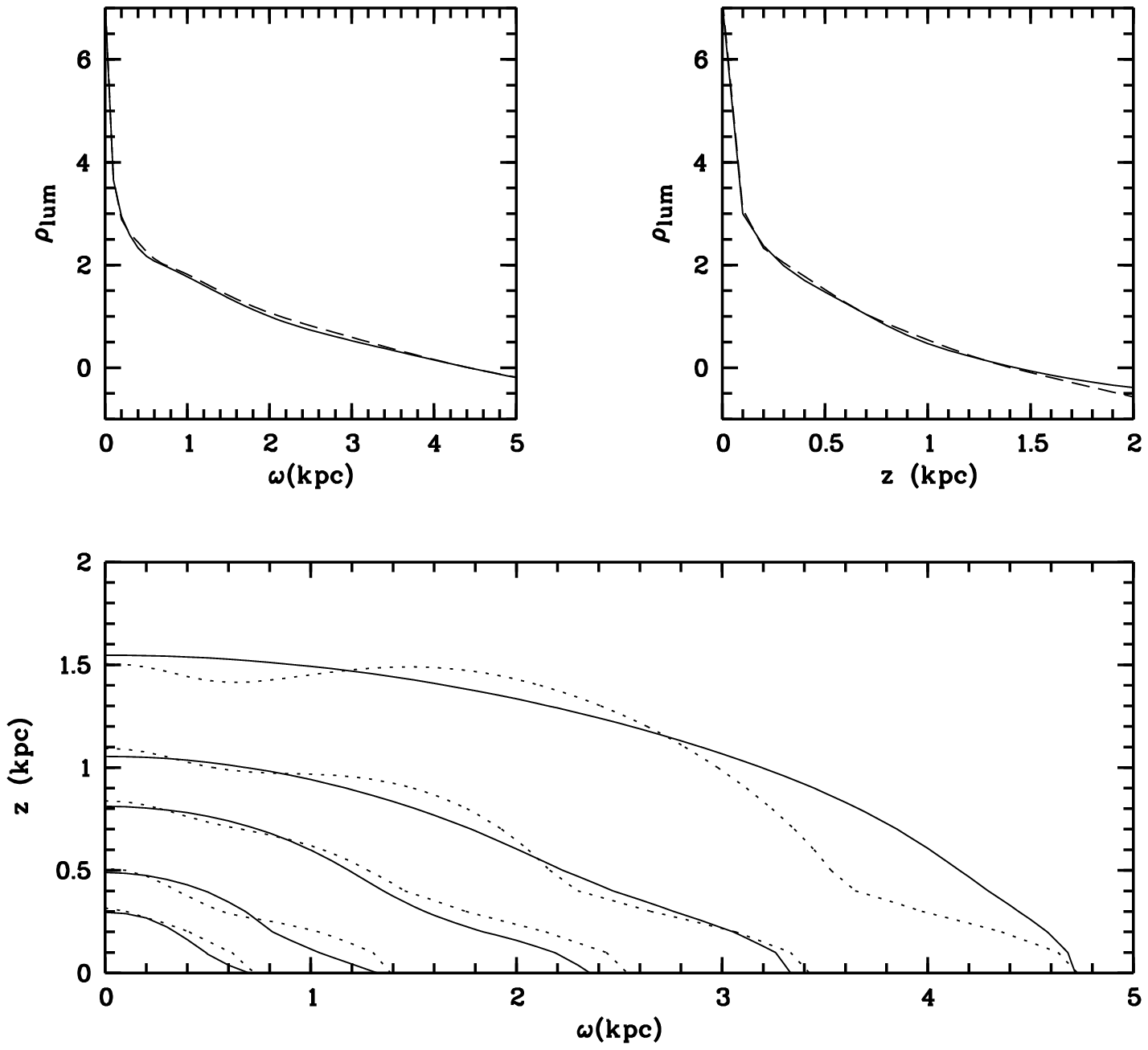,width=\hssize}
\caption[]{A comparison of the luminous mass density, as obtained by 
the MGE technique (solid lines), and the 3I fit (dotted or dashed
lines). The upper left panel shows the log of the luminous mass
density along the major axis, the upper right panel dito along the
minor axis }
\label{fig:showfitrho}
\figend
In Fig.~\ref{fig:compnobh-bh-centr} we compare two models on the major
axis, within the central $8''$. The solid line
corresponds to a model with a $6.5\times10^8$~\Msun black hole, the dashed
line has no black hole. Clearly, even in the three-integral case, the latter model fails
dismally, especially in the inner $2\arcsec$. 
This is to be expected of course: the large dispersion in
the centre exceeds the escape velocity if no black hole is present. Any
reshuffling of orbits in order to obtain a favourable viewing position
is insufficient in the absence of a black hole, because there is simply not
enough kinetic energy present. Although new components not included
in our library may help, this is a strong suggestion
that even three-integral models without a black hole cannot fit the data.

In fact, if we compactify integral space by considering the
representation in ($E$, $J\sqrt{E}$, $I_3 E$) then powers, such as
Fricke components, form a complete set, if one allows negative
coefficients. In practice, of course, one is limited by finite
computing resources. But at least, all orbits are present, with
relative weights that vary continuously, and thus not necessarily with
arbitrarily flexible relative weights. However, this practical
incompleteness is not fundamentally different from the one present in,
say, the Schwarzschild method. In the latter case, phase space cannot be
covered completely, also for practical reasons, but the orbits
considered can have arbitrarily flexible relative weights, at least
without the (necessary) smoothing.  So both methods are incomplete in
practice, be it to varying degree, but the incompleteness certainly
will manifest itself differently.
\figstart{figure=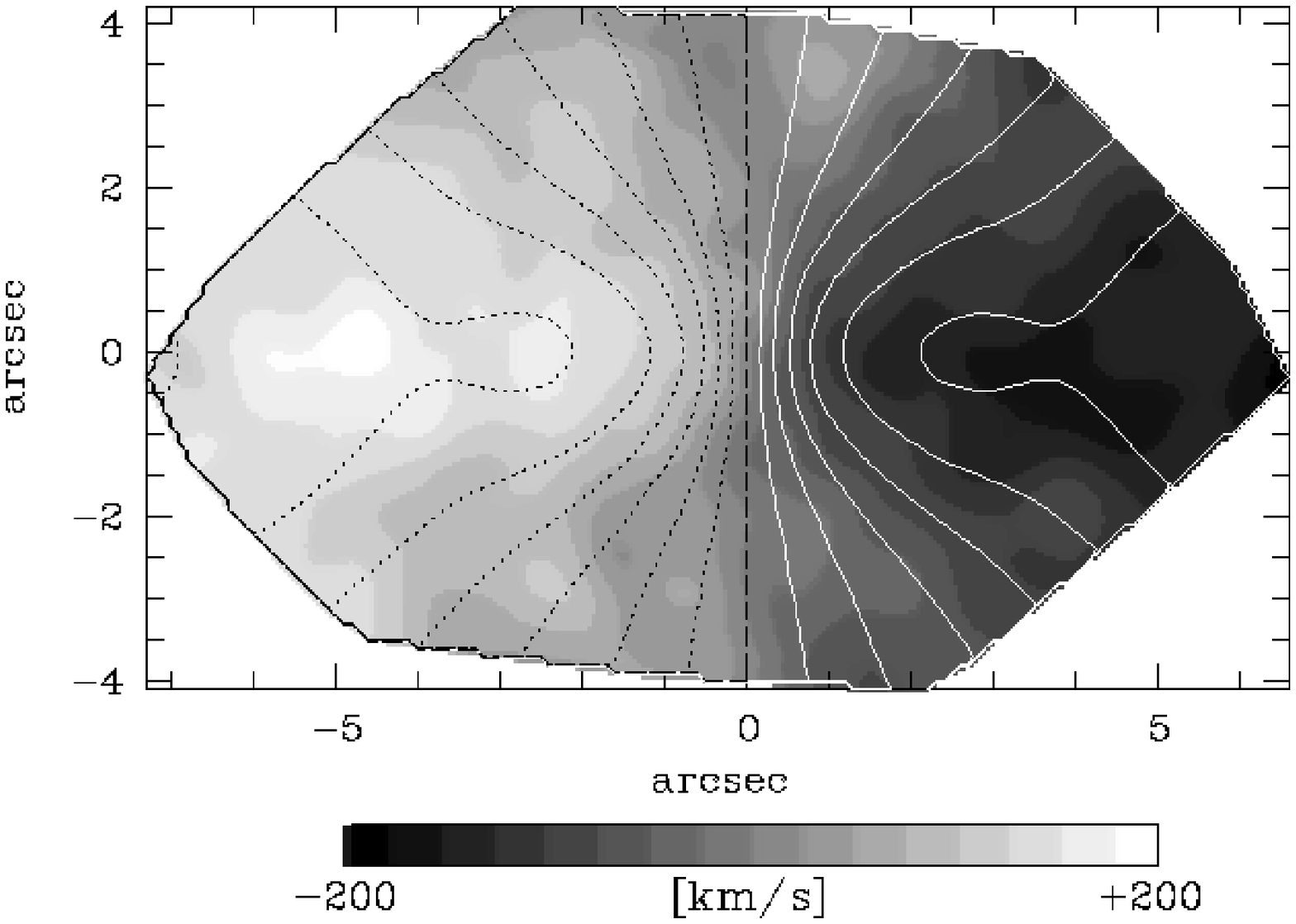,width=\hssize}
\psfig{figure=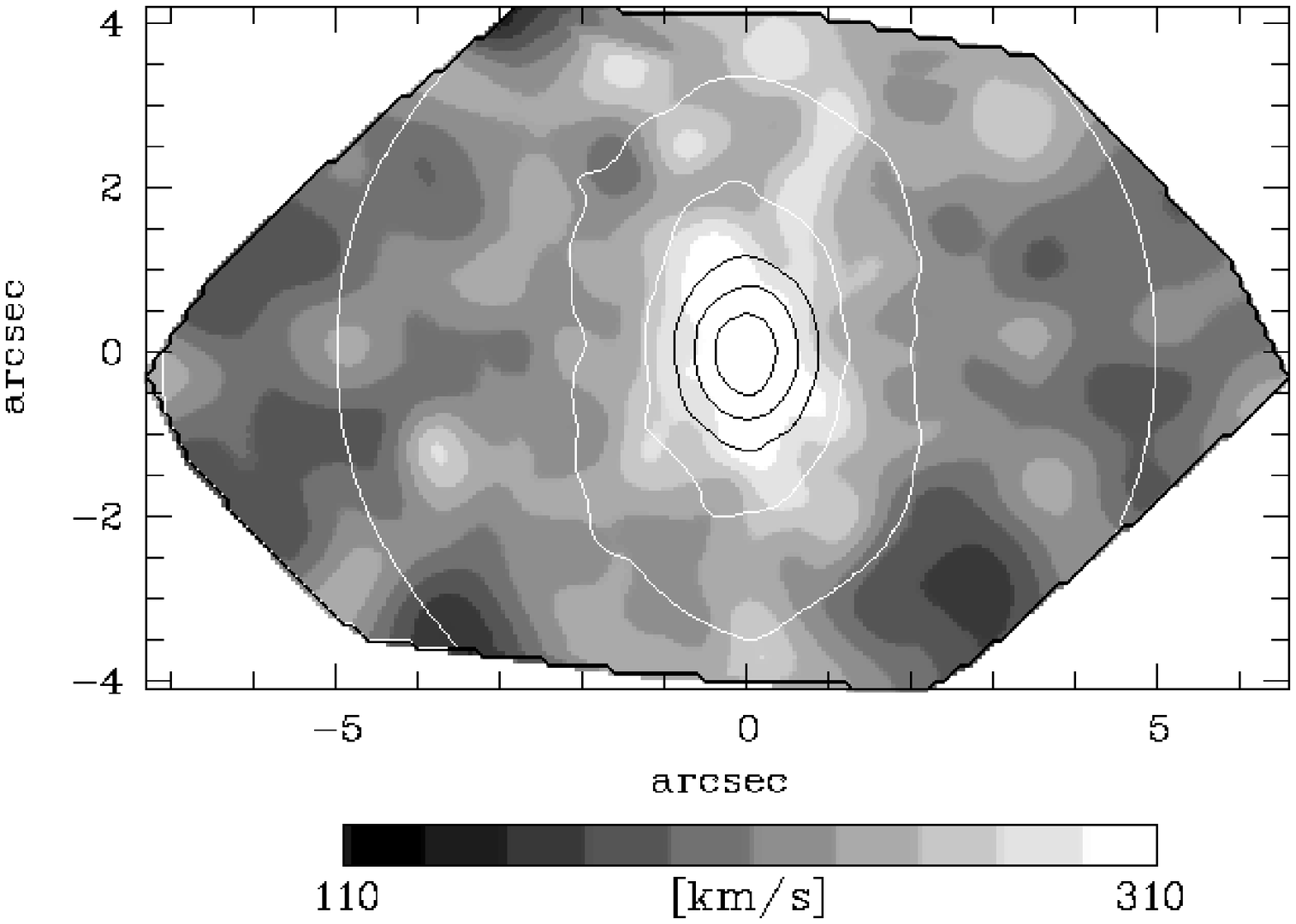,width=\hssize}
\caption[]{The TIGER maps of the first true velocity moments $\tilde{V}$ (upper panel) 
and $\tilde{\sigma}$ (lower panel), compared to the three-integral model with
$\MBH = 6.5\times10^8$~\Msun (contours: step of 20~\kms for $\tilde{V}$ and
25~\kms for $\tilde{\sigma}$).}
\label{fig:momtig_3I}
\figend

Regarding our second question, we first compare in
Fig.~\ref{fig:showfitrho} the luminous mass density as obtained with
the MGE technique (see Section \ref{sec:MGE}) with the best 3I model
that has a black hole of $6.5\times10^8$~\Msun (see also Fig.~\ref{fig:showcusp}). 
The fit is quite reasonable, certainly considering the dynamic range that is to be covered 
(see upper panels).

In Fig.~\ref{fig:momtig_3I}, \ref{fig:majaxis} and \ref{fig:offset} 
we next compare the input kinematical data with
our best model that has a black hole of $6.5\times10^8$~\Msun and includes
a dark halo. The fit is good, except perhaps at the centre along 
the cut $y=20\arcsec$ and the one $x=40\arcsec$ (Fig.~\ref{fig:offset}). 
In fact, the data (Illingworth \& Schechter 1982) presented in this plot
are mildly inconsistent with other published data
along the minor or major axis (e.g. vdM+94, see Fig.~\ref{fig:majaxis}).
The model with a constant mass to light ratio ($M/L_V = 10$, dotted lines
in Fig.~\ref{fig:majaxis} and \ref{fig:offset}) does however
show significant discrepancies along the major-axis and the cut
$y=20\arcsec$, with an overestimate of the velocity in the inner $50\arcsec$.
This shows, as already stated, that the mass to light ratio must significantly
increase outwards, by nearly a factor of 2 between the inner parts ($R < 50\arcsec$,
$M/L \sim 6$) and the last measured point ($R \sim 200\arcsec$, $M/L_V \sim 10$). 
We therefore confirm what was suggested by C+93 via simple dynamical
arguments, but here using general three-integral dynamical models.
Although a stellar mixture with a global $M/L_V \sim 10$ is not
excluded, it would have to be reconciled with the nearly constant (but decreasing) 
colours in the outer parts of NGC~3115 (for $R > 120\arcsec$,
Strom et al. 1977) and the metal poor
populations  observed in the halo by Elson (1997). The major-axis 
velocity and dispersion profiles are also more consistent with flat curves than with
falling ones (see Fig.~\ref{fig:majaxis} and Fig.~\ref{fig:offset}): this suggests
that the $M/L_V$ is still significantly increasing outwards for $R > 150\arcsec$.
In the view of these two points, we favour the hypothesis of the 
presence of a dark halo
to explain the observed kinematics of NGC~3115 at large radii.
\figstart{figure=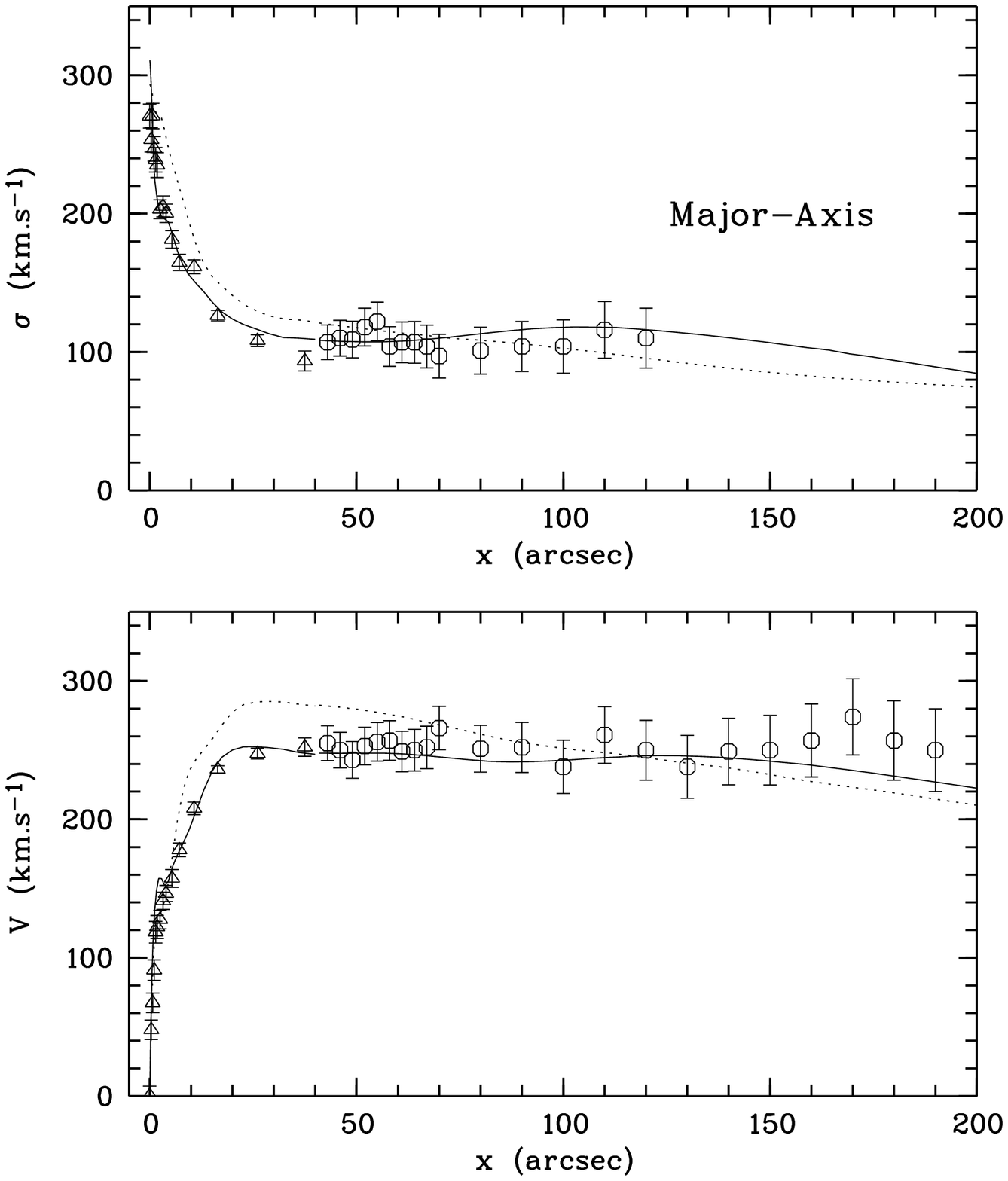,width=\hssize}
\caption[]{A comparison of the kinematics along the major-axis
and the best fits with a $6.5\times10^8$~\Msun black hole: including
a dark halo (solid lines) or with a constant $M/L_V = 10$ (dotted lines).
Inside 40 arcseconds we compare the model to the data of vdM+94, and outside this radius to 
the data of Capaccioli \etal (1993). The models have been convolved accordingly.}
\label{fig:majaxis}
\figend
\figstart{figure=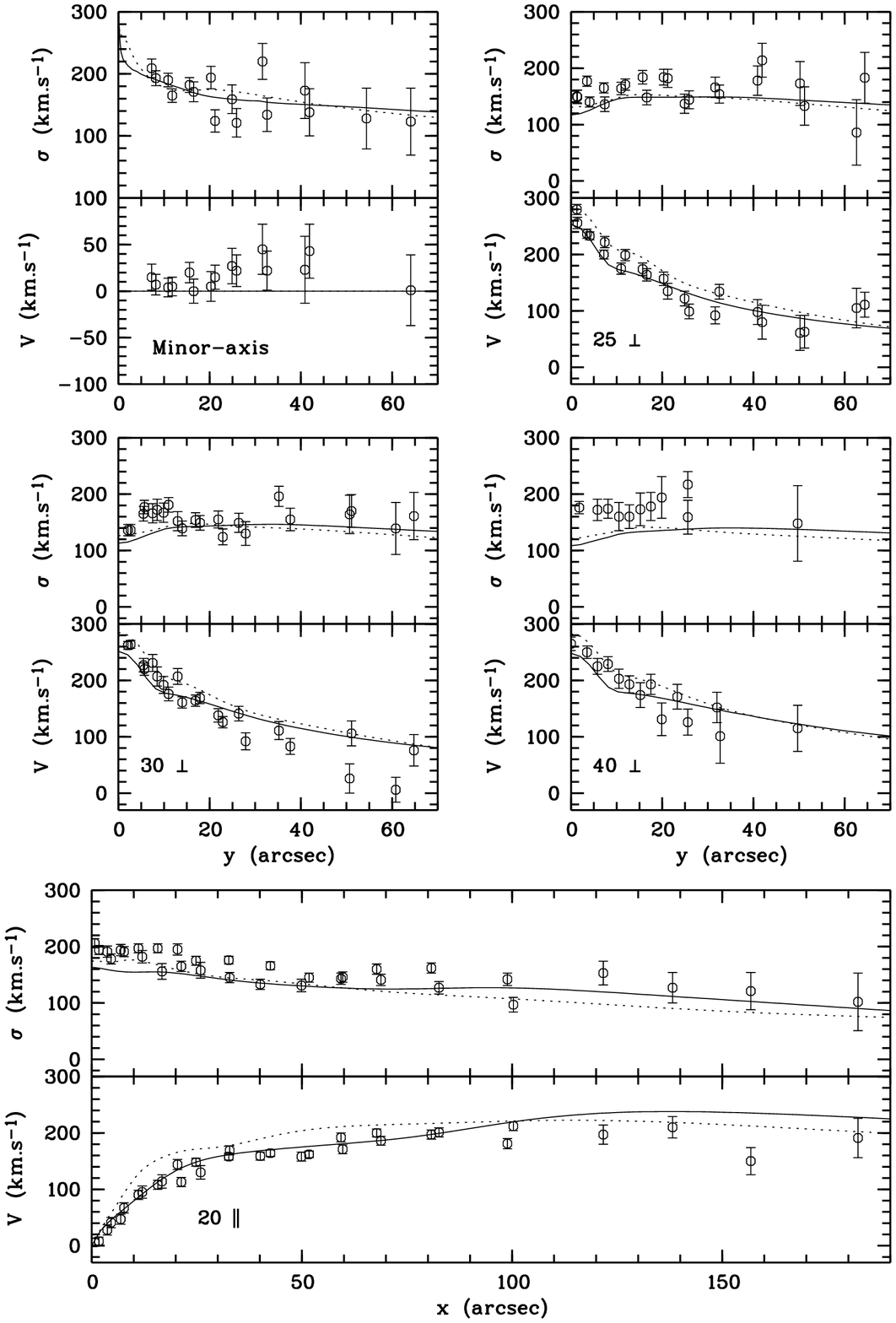,width=\hssize}
\caption[]{A comparison of the data points of Illingworth \& Schechter (1982) 
and the best fits with a $6.5\times10^8$~\Msun black hole: with a dark halo (see text,
solid lines) and with a constant $M/L_V = 10$ (dotted lines). For each cut, we present the
velocity (lower panel) and dispersion (upper panel). Note that the discrepancies
between the model with the dark halo (solid lines)
and the data are mostly due to inconsistencies in the data itself.}
\label{fig:offset}
\figend

For efficiency reasons, we included values for the hermite
coefficients at only 2 radii . At $7''$, the model convolved at the
resolution of vdM+94 data has $h_3=-0.0025$, close to the observed
$h_3=-0.05$, and in fact consistent with other published $h_3$ at this
radius (Bender et al. 1994, F96).  At $26''$ we obtain $h_3=-0.15$, a
much better fit than the 2I result, now perfectly consistent with the
observed $h_3=-0.16$. In fact all observed higher order moments are
now reasonably fit by the three-integral model.

Finally, we present in Fig.~\ref{fig:rotsigphi} and Fig.~\ref{fig:sigrz} the mean rotation
and velocity dispersions of the model in a meridian.
The inner disc is tangentially anisotropic with
$\sigma_{\phi} / \sigma_r$ reaching a value of $\sim 0.6$ in the inner
$2\arcsec$. This confirms the finding obtained via the two-integral model
in Sect.~\ref{sec:2I} that the inner disc seems to be close to maximum rotation. 
Between $3\arcsec$ and $15\arcsec$ in the equatorial plane,
the model is nearly isotropic. The inner spheroid ($R < 60\arcsec$) 
is close to two-integral and has a nearly constant $\sigma_{\phi} / \sigma_r \sim 0.65$. 
Going outwards, the distribution function becomes 
significantly three-integral and more radially anisotropic.
Already at a radius of $30\arcsec$ in the equatorial plane ($R \sim 1.5$~kpc) 
$\sigma_z / \sigma_r \sim 1.1$ and $\sigma_{\phi} / \sigma_r \sim 0.8$.
The ratio $\sigma_{\phi} / \sigma_r$ reaches a local maximum of $0.92$,
at a radius of $\sim 100\arcsec$, where the outer disc contribution
disappears and the dark matter becomes dominant.
\figstart{figure=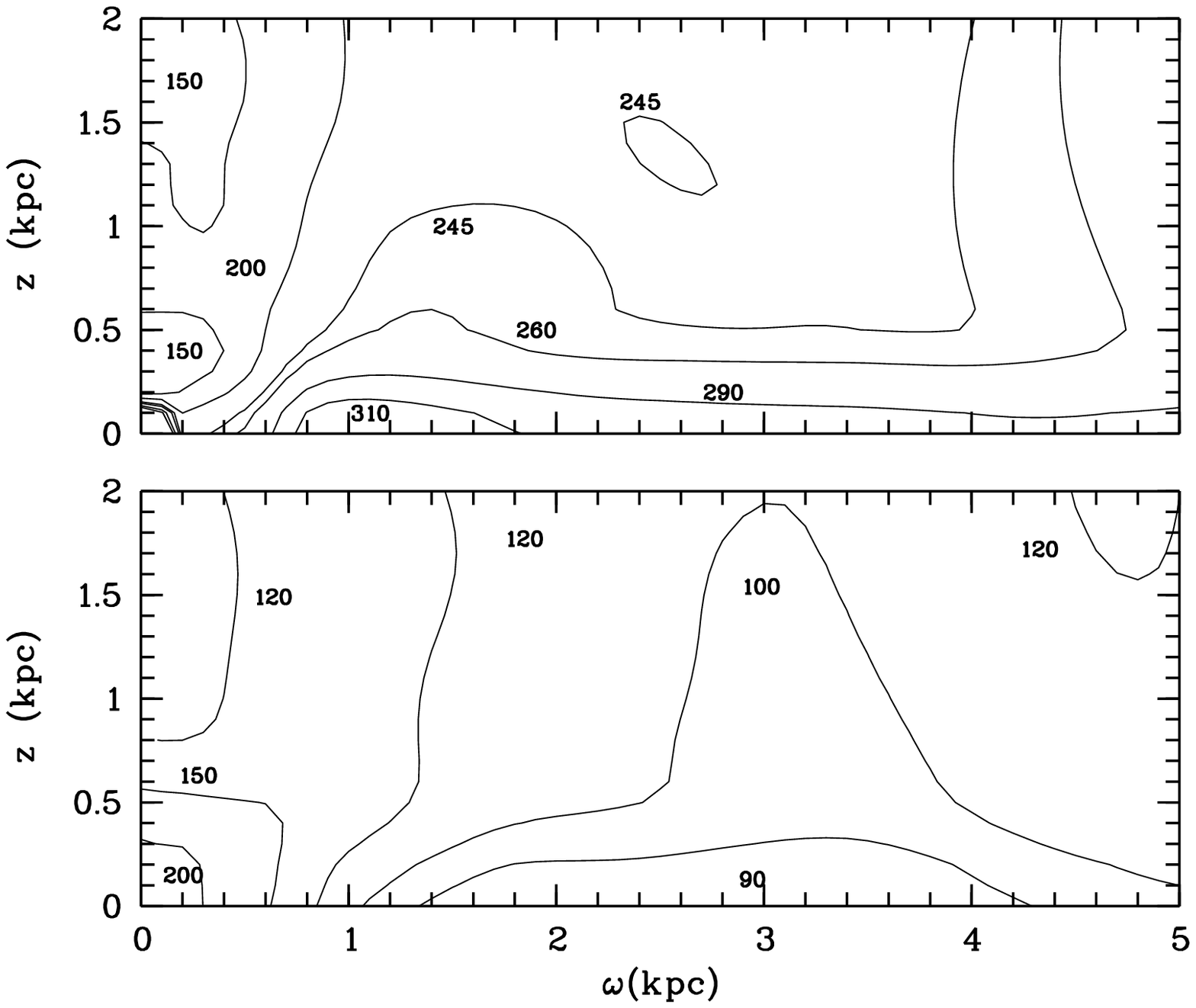,width=\hssize}
\caption[]{Some moments of the best fit 3I distribution function in a meridian.
Upper panel: $\langle v\rangle$ and lower panel: $\sigma_{\phi}$.}
\label{fig:rotsigphi}
\figend
\figstart{figure=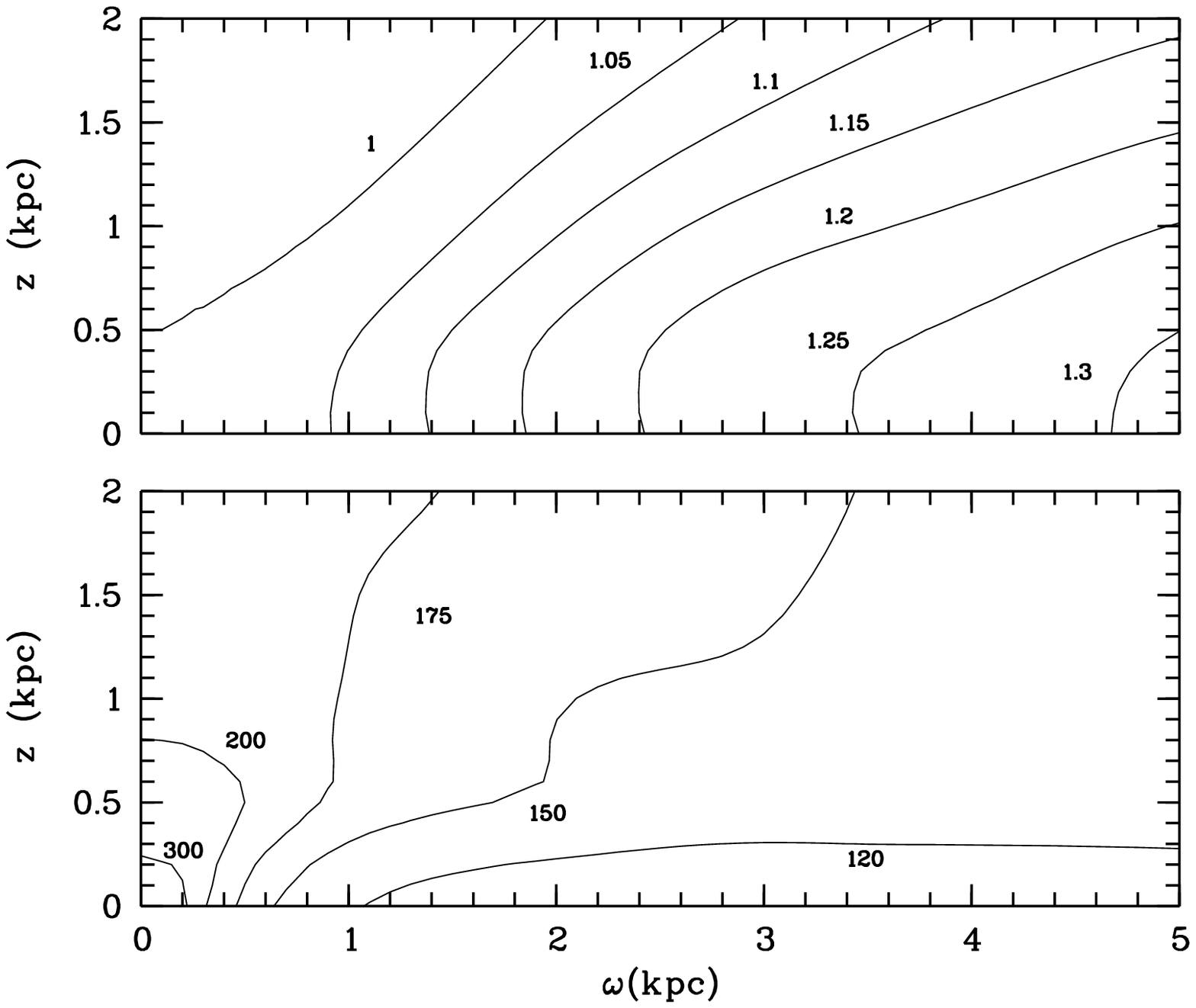,width=\hssize}
\caption[]{Some moments of the best fit 3I distribution function in a meridian.
Lower panel: $\sigma_r$ and upper panel: $\sigma_z / \sigma_r$.}
\label{fig:sigrz}
\figend

\section{Conclusions}
\label{sec:conc}

We have presented a detailed analysis of the kinematics
of NGC~3115 using different modeling techniques. As far as possible,
we have made use of the photometric and kinematical data available to us, 
in order to build dynamical models following realistic light and mass distributions.
For the first time, we have included two-dimensional spectroscopic data in order
to constrain the models in the central few hundred parcsecs.

Jeans equations were used to define reasonable ranges for the mass to light
ratio and the central dark mass in the frame of a two integral dynamical model.
Fitting the first two velocity moments in the central $40\arcsec$ gives
$\MBH = 6.5 \pm 3.5 \times 10^8$~\Msun\ and $M/L_V = 6.5 \pm 0.7$.
These estimations were then tested by retrieving the corresponding distribution functions
via the Hunter \& Qian method (Sect.~\ref{sec:2I}). 
We would like to emphasize the fact that a simple two-integral
model with a constant mass to light ratio 
can fit the velocity and dispersion profiles of NGC~3115 inside $\sim 45\arcsec$
surprisingly well (see Sect.~\ref{sec:2I}). 

However, no two-integral model could fit the value of $V$ 
and $h_3$ simultaneously\footnote{Remember that $h_3$ depends on $V$.}
as observed by vdM+94 around $25\arcsec$ along the major-axis, where the outer disc
significantly contributes to the surface brightness.
Models built using the quadratic programming technique
confirmed that three integral components solve
this problem (Sect.~\ref{sec:QP}). The dynamical structure 
of the outer part of NGC~3115 is therefore very probably three-integral.

In Sect.~\ref{sec:2I}, we showed that there are no two integral
axisymmetric models which can fit the kinematics of NGC~3115 in the central
region, without the addition of a central dark mass of about $10^9$~\Msun.
The best fit, which makes use of FOS/HST (Kormendy \etal 1996) 
and Gauss-Hermite moments, was obtained for $\MBH = 0.94\times10^9$~\Msun 
and $M/L_V = 6.1$.
This estimation of the central dark mass is at the upper limit
of the range derived from simple Jeans models.
This shows the need of high spatial resolution kinematics
including the higher order moments in order to more accurately
constrain the mass distribution.

The quadratic programming technique confirmed the need of a central
dark mass as even the three-integral models without a black hole
failed to fit the central rise in the velocity dispersion.  In
principle, one could object that, since a library of 3I distribution
functions is always limited, no matter what method is considered for
the modeling, this is no definite proof that a roughly constant $M/L$
model cannot fit the observables.  However such a model would have to
populate a lot of stars on radial orbits in the very centre. This
would produce a galaxy core that is extremely prone to radial orbit
instabilities, which, given the dynamical time of the order of
$8\times10^6$ years at a radius of $10\arcsec$ ($\sim 500$~pc), would
have relaxed into a more stable configuration a long time ago, or
would have produced a bar, evidence of which is not (yet) present.

Previous estimations of the central dark mass in NGC~3115
are surprisingly different from one paper
to the other. Kormendy \& Richstone (1992) derived a value of $\MBH$ from
1 to $2\times 10^9$~\Msun using spherical models corrected for flattening.
Then Kormendy et al. (1996), using the same formalism, 
confirmed the upper limit of $\sim 2\times 10^9$~\Msun by
including the SIS and HST/FOS data. Both papers used a mass to light ratio
of $M/L_V \sim 4$. Finally, Magorrian et al. (1998)
found a best fit value of $\MBH \sim 4.8\times10^8$~\Msun with
a $M/L_V \sim 6.75$ (both values scaled for $D = 10$~Mpc), noting
that their value for $\MBH$ was probably an underestimate.
This is consistent with the picture drawn in the present work
since we find that a value of $\MBH$ as low as $6.5 \times10^8$~\Msun 
(with a corresponding $M/L_V = 6.8$) is compatible with the existing data.
However, our best fit model has $\MBH = 0.94 \times10^9$~\Msun, and 
values as high as $\MBH = 2\times10^9$~\Msun
are excluded at more than the 3$\sigma$ level.
Our models, because they more precisely follow the observed 
light distribution and include the derivation of the full distribution
function, are improved estimations of the central dark mass
and mass to light ratio. 

We finally showed that the mass to light ratio
should increase by a factor of two between the inner parts
($R < 50\arcsec$) and the outer parts ($R > 100\arcsec$) of NGC~3115.
We produced a model which fits the flat rotation and dispersion profiles
along the major-axis and thus gave the first evaluation of the amount of dark matter
required to explain the observed kinematics at large radii:
in our model the dark matter represents about 50\% of the total mass inside 
$150\arcsec$ and nearly 70\% inside $300\arcsec$.
These values are obviously still uncertain as we did not
fully examine alternative spatial distributions or radial profiles. 

\section*{Acknowledgments}
EE wishes to thank Roeland van der Marel for deriving dynamical models
mentioned in the argument of Sect.~\ref{sec:2I}, for making unpublished
data available to us, and for stimulating discussions.
This work greatly benefited from a collaborative work with Eddie Qian
during his visit to the Leiden Sterrewacht (The Netherlands).

\appendix
\section{Convolution and pixel binning using a kernel}

It is possible to combine pixel integration and PSF smearing
in only two integrals (as otherwise would be four: two for the seeing
and two for the pixel integration). This is described in the following
in the case of circular or rectangular pixels (see also Qian \etal 1995)

\subsection{Circular pixel}

For a circular pixel (TIGER like lenses) of radius $d$, the observable 
$S_o$ at a position $(x', y')$ after convolution
with a single gaussian of dispersion $\sigma$ and integration on the pixel is:
\begin{eqnarray}
S_o (x',y') & = & \frac{1}{\pi d^2} \times \int_0^{\infty} r dr \int_0^{2 \pi} 
d\theta K(r) \nonumber \\
&& \hspace*{-0.5cm} S\left[x' + r \cos{\left(\theta + \theta_0\right)}, y' 
+ r \sin{\left(\theta + \theta_0\right)}\right]
\end{eqnarray}
where:
\begin{eqnarray}
K(r) = \int_0^{d / \sigma} s \; ds \; I_0\left(\frac{s r }{\sigma}\right)
e^{-\frac{sr}{\sigma}} e^{-\frac{1}{2} (s - r/\sigma)^2 }
\end{eqnarray}
Note that if the integrals are derived with fixed quadratures, the Kernel
has only to be calculated once, and only depends on the radius $r$.

\subsection{Rectangular pixel}

For a rectangular pixel of dimension $2l \times 2w$.
The observable $S_o$ at a position $(x', y')$ after convolution
with a single gaussian of dispersion $\sigma$ and integration on the pixel is:
\begin{eqnarray}
S_o (x',y') & = & \frac{1}{4 l w} \times \int_0^{\infty} r dr \int_0^{2 \pi} 
d\theta K(r, \theta) \nonumber \\
&& \hspace*{-0.5cm} S\left[x' + r \cos{\left(\theta + \theta_0\right)}, y' 
+ r \sin{\left(\theta + \theta_0\right)}\right]
\end{eqnarray}
where:
\begin{eqnarray}
K(r, \theta) & = & \frac{1}{4} \times 
\left[ {\rm erf}\left(\frac{l + r \cos{\theta}}{\sqrt{2}\sigma} \right)
- {\rm erf}\left(\frac{-l + r \cos{\theta}}{\sqrt{2} \sigma} \right)\right] 
\nonumber \\
&& \hspace*{-0.5cm} \times \left[{\rm erf}\left(\frac{w + r \sin{\theta}}{\sqrt{2} \sigma} \right)
- {\rm erf}\left(\frac{-w + r \sin{\theta}}{\sqrt{2} \sigma} \right)\right]
\end{eqnarray}
and note again that if the integrals are derived with fixed quadratures, the Kernel
can be calculated only once on a fixed grid.

\section{Cusps in MGE photometric models}

As we wish to build photometric models which can reproduce
central power laws, it has been necessary to extend the
MGE formalism (Emsellem et al. 1996) to include such cusp components. The adopted form
for the central component is then:
\begin{equation}
\nu(m_c^2) = I_{c} \cdot \left(\frac{m_c^2}{2 \cdot \sigma_c^2} \right)
^{- \gamma / 2} \times \exp{\left\{ - \frac{m_c^2}{2 \cdot \sigma_c^2} \right\}}
\end{equation}
where $m_c^2 = R^2 + Z^2 / q_c^2$.
This is the natural generalized form of the 3D gaussian function.

The integrated luminosity of this component is $I_c \cdot 2 \pi \left(\sqrt{2}
\sigma_c \right)^3 q_c \Gamma\left(1.5 - \gamma / 2\right)$, and the
gravitational potential contribution simply:
\begin{eqnarray}
\Phi(R, Z) & = & 4 \pi G q_c \sigma_c^2 P_c \times \\
&& \hspace{-2cm} \int_0^1 \frac{\Gamma\left(1 - \gamma
/ 2, \frac{T^2}{2 \cdot \sigma_c^2} \left(R^2 + Z^2 / \left(1 - e^2_c T^2 \right)
\right) \right)}{\left(1 - e^2_c T^2 \right)^{1/2}} \diff T
\end{eqnarray}
with $e^2_c = 1 - q^2_c$ and $P_c = {\cal M / L} \times I_c$.

\label{lastpage}

\begin{thebibliography}{99}
\bibitem{} Bacon R. et al., 1995, A\&AS, 113, 347
\bibitem{} Bacon R., Emsellem E., Monnet G., Nieto J.L., 1994, A\&A, 281, 691
\bibitem{} Batsleer P.,  Dejonghe H., 1995, A\&A, 294, 693
\bibitem{} Bender R., 1990, A\&A, 229, 441
\bibitem{} Bender R., Saglia R. P., Gerhard O., 1994, MNRAS, 269, 785
\bibitem{} Binney J., Mamon G.A., 1982, MNRAS, 200, 361
\bibitem{} Byun Y. I., Freeman K.C., 1995, ApJ, 448, 563 
\bibitem{} Capaccioli M., Cappelaro E., Held E. V., Vietri M., 1993, A\&A, 274, 69 (C+93)
\bibitem{} Capaccioli M., Held E. V., Nieto J.-L., 1987, AJ, 94, 1519
\bibitem{} Capaccioli M., Vietri M., Held E. V., 1988, MNRAS, 234, 335
\bibitem{} Cretton N., van den Bosch F., 1998, submitted to ApJ
\bibitem{} Dejonghe H., 1986, Phys. Rep., 133, 217
\bibitem{} Dejonghe H., 1987, ApJ, 320, 477
\bibitem{} Dejonghe H., 1989, ApJ, 343, 113
\bibitem{} Dejonghe H., De Bruyne, V., Vauterin, P., 1996, A\&A, 306, 363.
\bibitem{} de Vaucouleurs A., Longo G., 1988, Univ. of Texas, Austin
\bibitem{} Elson R. A. W., 1997, MNRAS, 284, 771 
\bibitem{} Emsellem E., 1995, A\&A, 303, 673
\bibitem{} Emsellem E., Monnet G., Bacon, R. 1994, A\&A, 285, 723
\bibitem{} Emsellem E., Bacon R., Monnet G. Poulain P., 1996, A\&A, 312, 777
\bibitem{} Fisher D., 1997, AJ, 113, 950 (F97)
\bibitem{} Fisher D., Franx M., Illingworth G., 1996, ApJ, 459, 110
\bibitem{} Hunter C., Qian E., 1993, MNRAS, 202, 401
\bibitem{} Illingworth G., Schechter P. L., 1982, ApJ, 256, 481
\bibitem{} Jarvis B. J., Freeman K. C., 1985, ApJ, 295, 314
\bibitem{} Kormendy J., Richstone D., 1992, ApJ, 393, 559 (KR92)
\bibitem{} Kormendy J., Bender R., Evans A., Richstone D., 1998, AJ, 115, 1823
\bibitem{} Kormendy J., Bender R., Richstone D., Ajhar E. A., Dressler A., Faber S. M.,
Gebhardt K., Grillmair C., Lauer T. L., Tremaine S., 1996, ApJL, 459, 57 (K+96)
\bibitem{} Magorrian J., Tremaine, S., Richstone D., Bender R., Bower G., Dressler A.,
Faber S. M., Gebhardt K., Green, R., Grillmair C., Kormendy J., Lauer T.,
1998, AJ, 115, 2285
\bibitem{} Monnet G., Bacon R., Emsellem E., 1992, A\&A, 253, 366
\bibitem{} Poulain P., 1986, A\&AS, 64, 225
\bibitem{} Poulain P., 1988, A\&AS, 72, 215
\bibitem{} Qian E. E., de Zeeuw P. T., van der Marel R. P., Hunter C., MNRAS, 1995,
274, 602
\bibitem{} Seifert W., Scorza C., 1996, A\&A, 310, 75
\bibitem{} Silva D. R., Boroson T. A., Thompson I. B., Jedrzejewski R. I., 1989, AJ, 98, 131
\bibitem{} Scorza C., Bender R., 1995, A\&A, 293, 20
\bibitem{} Strom K. M., Strom S. E., Jensen  E. B., Moller J., Thompson L. A., Thuan T. X.,
1977, ApJ, 212, 335
\bibitem{} van den Bosch F., de Zeeuw P. T., 1996, MNRAS, 283, 381
\bibitem{} van der Marel R. P., 1994, ApJL, 432, 91
\bibitem{} van der Marel R. P., Cretton N., de Zeeuw P. T., Rix H.-W., 1998, ApJ, 493, 613
\bibitem{} van der Marel R. P., de Zeeuw P. T., Rix H.-W., 1997, ApJ, 488, 119
\bibitem{} van der Marel R. P., Rix H.-W., Carter D., Franx M., White S. D. M., de Zeeuw T.,
1994, MNRAS, 268, 521 (vdM+94)
\bibitem{} Wagner S. J., Dettmar R.-J., Bender R., 1989, A\&A, 215, 243
\bibitem{} Worthey G., 1994, ApJS, 95, 107
\end{thebibliography}
\end{document}